\documentclass[a4paper,11pt]{article}
\pdfoutput=1 % if your are submitting a pdflatex (i.e. if you have
\usepackage{jheppub} % for details on the use of the package, please
\usepackage[T1]{fontenc} % if needed

\usepackage{graphicx}
\usepackage{slashed}
\usepackage{tabularx,ragged2e}
\usepackage{amssymb,fge}
\usepackage{amsmath,amssymb}
\usepackage{slashed}
\usepackage{dsfont}
\usepackage{verbatim}
\usepackage{subfig}
\usepackage{mathtools,ytableau,amsfonts,tikz}
\usepackage{graphicx}
\usepackage{physics}
\usepackage{amsthm}
\usepackage{tcolorbox}
\usepackage[normalem]{ulem}
\usepackage{mathrsfs}

\usepackage{xcolor}

\usepackage{scalerel,stackengine}
\stackMath
\newcommand\reallywidehat[1]{%
\savestack{\tmpbox}{\stretchto{%
  \scaleto{%
    \scalerel*[\widthof{\ensuremath{#1}}]{\kern-.6pt\bigwedge\kern-.6pt}%
    {\rule[-\textheight/2]{1ex}{\textheight}}%WIDTH-LIMITED BIG WEDGE
  }{\textheight}% 
}{0.5ex}}%
\stackon[1pt]{#1}{\tmpbox}%
}

\newcommand{\hodge}{{\star}}

\usepackage{tikz-cd}

\newcommand{\mybar}[3]{%
    \mathrlap{\hspace{#2}\overline{\scalebox{#1}[1]{\phantom{\ensuremath{#3}}}}}\ensuremath{#3}
}
\definecolor{Persianturquoise}{rgb}{0.4, 0.67, 0.63}
\newcommand{\FR}[1]{\textcolor{Persianturquoise}{[FR: #1]}}

\definecolor{azure(colorwheel)}{rgb}{0.0, 0.5, 1.0}
\newcommand{\Altytkn}[1]{\textcolor{azure(colorwheel)}{[AE: #1]}}

\newcommand{\toghether}[1]{\textcolor{blue}{[BOTH: #1]}}

\title{Unimodular time in JT gravity: a holographic clock}

\author[a]{Altay Etkin}%\note{Corresponding author.}}
\author[b]{and Farbod-Sayyed Rassouli}

\affiliation[a]{School of Mathematical Sciences and STAG Research Centre, University of Southampton,\\Highfield Campus, Southampton, SO17 1BJ, United Kingdom}
\affiliation[b]{School of Physics and Astronomy and Nottingham Centre of Gravity, University of Nottingham,\\University Park, Nottingham, NG7 2RD, United Kingdom}

\emailAdd{a.etkin@soton.ac.uk}
\emailAdd{sayyed.rassouli@nottingham.ac.uk}

\abstract{How is a ``bulk clock'' encoded holographically? We address this in Jackiw-Teitelboim (JT) gravity, where a natural physical clock emerges by promoting the vacuum energy to a dynamical variable: the vacuum cosmological constant becomes a top form degree of freedom conjugate to spacetime volume, thereby defining a notion of bulk physical time. This construction is naturally formulated in the Henneaux-Teitelboim (HT) framework. We show that the boundary dynamics is the Schwarzian mode coupled to a free particle on $U(1)$, matching the universal low-energy effective action of the complex SYK model. By further clarifying the role of the vacuum cosmological constant as a top form, we establish the equivalence between JT gravity coupled to two-dimensional Maxwell theory and 2d HT gravity via an explicit field redefinition. The initial question is addressed: we show that the resulting boundary theory can itself be rewritten as an observer action, equivalently a $(0+1)$-dimensional HT theory. This yields a direct identification of the boundary clock with the $U(1)$ phase mode, and makes its relation to the bulk clock explicit.}

\begin{document} 
\maketitle
\flushbottom

%--------------------------------------------------------------------
\section{Introduction and summary}

In physics, time is what a clock measures, and in the absence of gravity, time evolution is generated by the Hamiltonian. In gravity, the situation is subtler. The Wheeler-DeWitt constraint makes the theory appear ``timeless'', so one is naturally led to look for an internal clock, a physical degree of freedom whose conjugate momentum can serve as a generator of evolution. A simple and robust candidate is the accumulated spacetime volume. Since the cosmological constant multiplies the spacetime volume in the action, it is natural to ask whether one can treat the cosmological constant as a dynamical variable and interpret the quantity conjugate to it as a physical time parameter. This idea is realised cleanly in the Henneaux-Teitelboim (HT) formulation of unimodular gravity \cite{Henneaux:1989zc}, where the cosmological constant becomes a variable and a corresponding unimodular time emerges as its canonical conjugate \cite{unimod1,Bombelli:1990ke,Kuchar:1991xd}. Within this framework, the vacuum cosmological constant can be viewed as a simplified model of an observer \cite{Witten:2023xze,Alexandre:2025rgx,Blommaert:2025bgd,Blommaert:2025rgw}. 

A concrete working model of such a physical clock also arises in two-dimensional JT gravity. In particular, a gauge-theoretic formulation of JT gravity with a vacuum cosmological constant naturally involves a central extension of the relevant isometry algebra, and this extension organises an additional sector in which the cosmological constant is treated as a dynamical variable with a canonical conjugate \cite{Cangemi:1993bb,Alexandre:2025rgx}. In this formulation, the would-be ``frozen'' Hamiltonian constraint admits a Schr\"{o}dinger-like evolution with respect to the conjugate variable, providing a controlled notion of bulk physical time in a setting where time is otherwise subtle \cite{Alexandre:2025rgx}. 

This two-dimensional construction provides a clean starting point for the broader question pursued in this work: what is its holographic description, and how is this bulk clock encoded at the boundary?

\paragraph{Background and motivation.}

An important viewpoint often missed is that the cosmological constant can be interpreted as the effective contribution of a top form field strength. In higher-dimensional constructions, this idea appears naturally. For instance, in four dimensions, the vacuum energy can be tied to a four-form flux \cite{Aurilia:1980xj,Hawking:1984hk,Weinberg:1988cp,Brown:1987dd,Duff:1989ah}, while in supergravity theories, the same logic is ubiquitous in flux compactifications \cite{Freund:1980xh,Duff:1986hr,Douglas:2006es} and in the effective descriptions of warped throats that appear in near-horizon analysis \cite{Maldacena:1997re,Witten:1998qj,Aharony:1999ti}. In such settings, the cosmological constant is often better thought of as an emergent quantity controlled by higher-form fields rather than a rigid coupling.

%--------------------------------------------------------------------------------------

This aspect of the cosmological constant, expressed in terms of top forms, is relevant within holographic theories. In higher dimensions, this viewpoint is ubiquitous in the near-horizon physics of near-extremal black holes, where an $AdS$ region supported by a top form field strength emerges. A closely analogous phenomenon occurs in two dimensions, as near-extremal four-dimensional black holes develop an $AdS_2 \times S^2$ throat, whose spherically reduced dynamics are captured by JT gravity \cite{Jackiw:1984je,Teitelboim:1983ux,Maldacena:2016upp,Jensen:2016pah,Engelsoy:2016xyb} coupled to top form fields \cite{Bardeen:1999px,Almheiri:2016fws,Nayak:2018qej,Moitra:2019bub}. The resulting two-dimensional holographic setup is particularly fertile because it is solvable enough: JT gravity isolates the boundary reparametrisation mode governed by the Schwarzian action \cite{Almheiri:2014cka,Maldacena:2016upp,Jensen:2016pah,Engelsoy:2016xyb}, and this same boundary dynamics arises as the universal low-energy sector of the SYK model. Taken together, these facts make $AdS_2$ an ideal laboratory to study holography through the lens of HT gravity, since it packages the vacuum cosmological constant, the relevant top form sector, and the associated notion of unimodular time into a single holographic framework.

%--------------------------------------------------------------------------------------

The main motivation for this work is that, despite the conceptual appeal of HT gravity, its holographic interpretation has remained largely unexplored in the literature. This gap is especially striking given a recent realisation in two-dimensions, where holography can be explicitly implemented as aforementioned. In other words, two-dimensional gravity provides a bridge between HT gravity and holography explicitly. On the other hand, a satisfactory notion of bulk and boundary physical time in JT gravity is still missing from the literature, despite recent attempts to incorporate observers as point-particles in JT gravity \cite{Abdalla:2025gzn,Blommaert:2025bgd,Blommaert:2025rgw}.

In this paper, we initiate the first holographic analysis of HT gravity by focussing on its Euclidean formulation in $AdS_2$. The Euclidean perspective is not merely technical; it sharpens the role of boundary conditions, clarifies the emergence of the boundary degrees of freedom, and makes the unimodular sector particularly transparent. 

\subsection{Summary of results}

The main findings of this work are:
\begin{itemize}
    \item We show that 2d Euclidean HT gravity in $AdS_2$, with mixed boundary conditions, is holographically captured by the same universal boundary dynamics as the low-energy effective theory of complex SYK: the Schwarzian mode together with a free particle on $U(1)$. 
    \item The vacuum cosmological constant admits a natural description in terms of top forms and is realised by 2d Maxwell theory. As a consequence, we show explicitly that JT gravity coupled to 2d Maxwell theory can be rewritten as HT gravity. In particular, once the appropriate boundary terms are included, the two formulations yield the same particle on $U(1)$ boundary action.
    \item At the boundary, we obtain a $(0+1)$-dimensional HT theory. This provides another realisation of the particle on a group and allows one to interpret the effective complex SYK boundary dynamics as $(0+1)$-dimensional HT unimodular gravity. Crucially, this view is complementary with the notion of a boundary observer action, where the Schwarzian couples to a boundary clock given by the phase field, which is shown to be the boundary unimodular time. Its relation to the bulk unimodular time is also discussed.
\end{itemize}

\paragraph{A two-dimensional HT gravity theory.}

In four dimensions, General Relativity (GR) admits a volume-preserving diffeomorphism-invariant formulation, commonly referred to as unimodular gravity \cite{Einstein-unimod,Weinberg:1988cp,Henneaux:1989zc,unimod1,Smolin:2009ti,Smolin:2010iq,Kaloper:2013zca,Padilla:2014yea,Bufalo:2015wda,Nemanja:2016sstft,Percacci:2017fsy,lombriser2019cosmological,Carballo-Rubio:2022ofy}. This formulation is classically equivalent to GR and is achieved by imposing the gauge fixing condition $\sqrt{-g} = 1$ (for more details about the original formulation of unimodular gravity, see appendix \ref{app:A}). The seminal work of Henneaux and Teitelboim \cite{Henneaux:1989zc} extended this framework by restoring full diffeomorphism invariance in unimodular gravity by promoting the vacuum cosmological constant $\Lambda$ to an off-shell dynamical variable. Other extensions of this theory to incorporate other constants of nature and make them dynamical variables have been explored extensively in the literature, see \cite{Alexander:2018djy,Magueijo:2020ntm,Magueijo:2021rpi,Magueijo:2021pvq,Jirousek:2018ago,Jirousek:2020vhy,Vikman:2021god,Etkin:2023amf}. This parametrisation becomes evident when the unimodular condition is modified to $\sqrt{-g} = \partial_\mu \mathcal{T}^\mu$, treated as an on-shell expression, with Lorentzian action given by
\begin{equation}\label{eq:HTgravityaction}
    S = \frac{1}{2} \int_{\mathcal{M}} d^4x \, \sqrt{-g}\,(R-2 \Lambda) + \int_{\mathcal{M}} d^4x \, \Lambda \partial_\mu \mathcal{T}^\mu,
\end{equation}
where $\mathcal{T}^\mu$ is an auxiliary vector density, and $\Lambda$ is promoted to an off-shell variable, with the condition that it is constant on-shell. Hence, this theory yields equations of motion identical to those given by the standard Einstein–Hilbert action.

While HT gravity is most commonly discussed in four dimensions, it also often arises implicitly in two-dimensional settings. In particular, several formulations of JT gravity, including those developed in the context of flat holography \cite{Gonzalez:2018enk,Afshar:2019tvp,Afshar:2019axx,Godet:2021cdl,Afshar:2021qvi,Kar:2022sdc,Rosso:2022tsv,Afshar:2022mkf}, can be understood as instances of HT gravity \cite{Callan:1992rs,Cangemi:1992bj,Jackiw:1992ev,Cangemi:1992ri,Grignani:1992hw,Cangemi:1992up,Kim:1992fb,Kim:1992ht,Cangemi:1993sd,Cangemi:1993bb,Jackiw:1993gf,Godet:2020xpk}.

A two dimensional HT gravity (HT$_2$) theory has been formulated \cite{Alexandre:2025rgx} based on JT gravity, with Lorentzian action given by:
\begin{equation}\label{eq:HT2 action}
S_{\text{HT}_2} :=\frac{1}{2} \int_{\mathcal{M}} d^2x \, \sqrt{-g} \,\Bigl(\phi\bigl(R -2\lambda\bigr)-2\Lambda\Bigr) + \int_{\mathcal{M}} d^2x \, \Lambda \partial_\mu \mathcal{T}^\mu,
\end{equation}
where $\lambda$ determines the global spacetime geometry on-shell, $\Lambda$ serves as the vacuum energy scalar field, and $\mathcal{T}^{\mu} = \varepsilon^{\mu \nu} A_\nu$ is the densitised dual of the $U(1)$ gauge field $A_\mu$. The variation of the action with respect to $\phi$, $g_{\mu\nu}$, $\mathcal{T}^\mu$, and $\Lambda$, yields the following equations of motion: 
\begin{align}
    R &= 2\lambda, \qquad \qquad \bigl(\nabla_{\mu} \partial_{\nu} - g_{\mu \nu} \Box\bigr)\phi = (\lambda\phi+\Lambda) g_{\mu \nu}, \nonumber\\
    0 &= \partial_\mu \Lambda, \qquad \qquad \sqrt{-g} = \partial_\mu \mathcal{T}^\mu.
\end{align}
From these, the on-shell dynamics of HT$_2$ gravity matches those of the JT model, provided that the unimodular HT constraint $\sqrt{-g} = \partial_\mu \mathcal{T}^\mu$ is satisfied, and that the vacuum cosmological constant is constant on-shell. Therefore, the classical dynamics remain unchanged and they describe the same physics.

Assuming an appropriate slicing of the Lorentzian manifold as $\mathcal{M} \cong \Sigma \times \mathbb{R}$, with constant-time spacelike hypersurfaces $\Sigma$, the unimodular constraint can be integrated as follows:
\begin{equation}
    \int_{\Sigma_0}^{\Sigma_t} d^2x \, \sqrt{-g}=\int_{\Sigma_t} d\Sigma \,  \mathcal{T}^0 - \int_{\Sigma_0} d\Sigma \, \mathcal{T}^0:=\Delta T_{\Lambda}.
\end{equation}
This gives us a definition for unimodular time on a slice $\Sigma$:
\begin{equation}
T_{\Lambda}(\Sigma):=\int_{\Sigma} d\Sigma \,  \mathcal{T}^0 = -\int_{\Sigma} d\Sigma \ A_1,
\end{equation}
with $\mathcal{T}^0 = -A_1$. This definition of unimodular time guarantees that it represents a physical flow of time. Identifying $\Delta T_{\Lambda}$ with the spacetime volume lying in the past of the leaf $\Sigma_t$, measured relative to a conventional initial leaf $\Sigma_0$, one finds the on-shell relation between spacetime volume and unimodular time:
\begin{align}\label{eq:deltatime}
    \Delta T_{\Lambda}:= T_{\Lambda}(\Sigma_t) - T_{\Lambda}(\Sigma_0) = \mathrm{Vol}(\Sigma_t \rightarrow \Sigma_0).
\end{align}
Because the spacetime volume swept out between the two hypersurfaces is strictly positive, $T_{\Lambda}$ is necessarily monotonic, assuming non-singular manifolds. Therefore, this notion of unimodular time not only introduces a physical clock but also incorporates physical observers naturally in gravity \cite{Witten:2023xze,Blommaert:2025bgd,Blommaert:2025rgw}. 

From the canonical Hamiltonian perspective \cite{Henneaux:1989zc,unimod1,Bombelli:1990ke,Kuchar:1991xd,Bombelli:1991jj,Daughton:1993uy,sorkin1,Daughton:1998aa,Smolin:2009ti,Smolin:2010iq,Magueijo:2021rpi,Magueijo:2021pvq,Isichei:2022uzl, Etkin:2023amf,Gielen:2024lpm,Gielen:2025ovv,Gielen:2025grd,Hallam:2025hdb}, the vacuum cosmological constant becomes canonically conjugate to unimodular time, with Poisson bracket $\{T_\Lambda,\Lambda\}_{\text{P.B.}}=1$. Upon canonical quantisation, this replaces the usual Hamiltonian constraint and Wheeler-DeWitt equation by a Schr\"{o}dinger-type evolution equation in unimodular time \cite{Alexandre:2025rgx}. A simple counting of constraints shows that the theory has no new local degrees of freedom, justifying the appearance of the topological term $\partial_{\mu} \mathcal{T}^{\mu} = \varepsilon^{\mu \nu} \partial_\mu A_\nu$. In addition to the fact that the theory is diffeomorphism invariant, there is an additional $\text{U}(1)$ gauge symmetry of the $\text{HT}_2$ action. Under $\mathcal{T}^{\mu} \rightarrow \mathcal{T}^{\mu} + \xi^{\mu}$ with $\partial_{\mu} \xi^{\mu} = 0$, there is one gauge degree of freedom left.

As a final comment, flat holography in two-dimensional dilaton gravity with a vanishing cosmological constant $\lambda = 0$, appears most naturally in the Cangemi–Jackiw model \cite{Cangemi:1992bj}---often written as $\widehat{\text{CGHS}}$ gravity \cite{Afshar:2019axx,Godet:2021cdl,Kar:2022sdc,Rosso:2022tsv,Afshar:2022mkf,Kar:2022vqy}. Recent work has shown that this theory is dual to a particular sector of the low-energy effective action of the complex SYK model, which is not surprising since the $\reallywidehat{\text{CGHS}}$ model is flat HT$_2$ gravity
\cite{Alexandre:2025rgx}. Furthermore, flat HT$_2$ has been demonstrated to capture a universal sector of near-horizon perturbations of nearly non-extremal black holes in higher dimensions \cite{Godet:2021cdl}.

\subsection{Outline of the paper}

This paper is organised as follows.

In \textbf{section} \ref{sec:2}, we begin by presenting the Euclidean action of HT$_2$ gravity together with the boundary terms required for a well-posed variational principle. Imposing mixed boundary conditions, we evaluate the on-shell action and show that the boundary dynamics reduce to the Schwarzian theory coupled to a particle on $U(1)$. We then extend the analysis to the thermal black-hole setting and conclude with a brief discussion of the thermoelectric properties of the resulting boundary theory. Throughout the section, we emphasise the boundary dynamics itself, postponing a detailed discussion of unimodular time to later.

In \textbf{section} \ref{sec:3}, we emphasise the role of top forms and their relation to the vacuum cosmological constant, and show that JT gravity coupled to 2d Maxwell theory can be recast as HT$_2$ gravity via an appropriate field redefinition, keeping careful track of the boundary terms on both sides. We also discuss how gauge transformations are realised under this change of variables. We then switch gears to a first order formulation, where HT$_2$ admits a gauge theoretic description as a BF theory with the gauge group given by a centrally extended $AdS_2$ isometry algebra. Using the triviality of this central extension, we show that the theory can be written in a decoupled form consisting of JT gravity together with a $U(1)$ BF sector. Finally, we comment on how HT$_2$ emerges from the near-horizon region of higher-dimensional black holes.

In \textbf{section} \ref{sec:4}, building on the field-redefinition technology developed in the previous section, we show that the boundary theory describing the low-energy effective dynamics of complex SYK can be rewritten as a $(0+1)$-dimensional HT theory. This reformulation naturally possesses a boundary unimodular time, which is identified with the phase field and makes it apparent that the Schwarzian mode is coupled to an observer action. In this language, the relation between bulk and boundary volumes becomes particularly transparent, and it sharpens the connection between the chemical potential and the boundary vacuum cosmological constant. We conclude the section with a discussion of the choice of abelian gauge group, $U(1)$ versus $\mathbb{R}$, and explain why working with either one does not affect the HT sector, and therefore leaves our main conclusions unchanged.

%----------------------------------------------------------------------------------

\section{Euclidean HT\texorpdfstring{$_2$}{HT2} gravity and complex SYK realisation}\label{sec:2}

In this section, we study Euclidean HT$_2$ gravity and derive its boundary dynamics, explicitly showing how the effective action of complex SYK emerges. We begin by fixing our notation and conventions. Our presentation is intentionally minimalistic: we will not invoke any of the unimodular machinery here, so as to study the theory on its own merits. The unimodular features and their implications will be taken up separately in the following sections. 

Consider the bulk Lorentzian action for HT$_2$ gravity with negative cosmological constant $\lambda = -1$\footnote{Throughout the paper, we shall use units in which the $AdS_2$ radius is set to $\ell = 1$. Restoring it, would give $\lambda = -\ell^{-2}$.}:
\begin{equation}
    S_{\text{bulk}} = \frac{1}{2} \int_{\mathcal{M}} d^2x \, \sqrt{-g}\bigl(\phi(R +2) - 2 \Lambda\bigr) + \int_{\mathcal{M}} d^2x \, \Lambda \,\partial_{\mu} \mathcal{T}^{\mu}.
\end{equation}
To get the corresponding Euclidean action, we perform a Wick rotation in coordinate time $t = -i \tau$. Demanding reality of the Euclidean action $I = -i S$ and the associated fields, we find the bulk Euclidean action to be:
\begin{equation}\label{eq:bulkHT2action}
    I_{\text{bulk}} = -\frac{1}{2} \int_{\mathcal{M}} d^2x \, \sqrt{g}\bigl(\phi(R +2) - 2 \Lambda\bigr) - \int_{\mathcal{M}} d^2x \, \Lambda \,\partial_{\mu} \mathcal{T}^{\mu}.
\end{equation}

Hence, the Euclidean HT$_2$ action is JT gravity supplemented by the unimodular HT part. Half of the equations of motion for HT$_2$ gravity are equivalent to those of JT gravity, that is
\begin{equation}
    R + 2 = 0, \qquad \qquad (\nabla_{\mu} \partial_{\nu} - g_{\mu \nu} \, \Box ) \phi = (\phi - \Lambda) g_{\mu \nu}
\end{equation}
The other half consists of the on-shell constancy of the vacuum cosmological constant and the unimodular HT condition, respectively:
\begin{equation}\label{eq:HTconstraint}
    \partial_\mu\Lambda = 0, \qquad \qquad \sqrt{g} = \varepsilon^{\mu \nu} \partial_{\mu} A_{\nu}.
\end{equation}
To treat HT$_2$ gravity holographically, one must first solve these equations of motion and then extract the asymptotic falloffs of the respective fields, thereby determining their boundary behaviour.

\paragraph{JT gravity and the Schwarzian.}
Since $R=-2$, is satisfied, we have global Euclidean $AdS_2$ metrics. Solving the equations of motion in Poincar\'{e} coordinates $(t,z)$, we find that the metric satisfies
\begin{equation}
    ds^2 = \frac{dt^2 + dz^2}{z^2}.
\end{equation}
The dilaton solution in these coordinates reads
\begin{equation}
    \phi(t,z) = \frac{a + b t + c(t^2 + z^2)}{z} + \Lambda_0,
\end{equation}
where $\Lambda_0$ is the on-shell constant vacuum cosmological constant. The asymptotic $AdS_2$ boundary conditions on the metric $g_{\mu \nu}$ and dilaton $\phi$ are such that
\begin{equation}
    ds^2\big|_{\partial \mathcal{M}} = \frac{du^2}{\epsilon^2}, \qquad \qquad \phi\big|_{\partial \mathcal{M}} \equiv \phi_b= \frac{\overline{\phi}_r}{\epsilon} + \Lambda_b.
\end{equation}
Here, $u$ is the coordinate on the boundary curve, $\overline{\phi}_r$ is the renormalised value of the dilaton taken to be a constant, and $\phi_b$ is the boundary value of the dilaton \cite{Maldacena:2016upp}. Note that we have introduced a cut-off $\epsilon$, which tends to zero as one approaches the $AdS_2$ boundary located at $z=0$. It is simple to check that upon parametrising the boundary curve by $(t(u),z(u))$, the metric can be rewritten as
\begin{equation}
    ds^2 = \frac{t'^2 + z'^2}{z^2} du^2 \equiv g_{uu} du^2 \, \, \, \, \Longrightarrow \, \, \, \, g_{uu}\big|_{\partial \mathcal{M}} \equiv h_{uu} = \frac{1}{\epsilon^2}.
\end{equation}
Imposing boundary conditions yields the perturbative result that $z = \epsilon t' +\mathcal{O}(\epsilon^3)$. In fact, the induced metric $h_{uu}$ is precisely the boundary value of $g_{uu}$, with determinant $h \equiv \mathrm{det}(h_{uu}) = \tfrac{1}{\epsilon^2}$.

\paragraph{Topological $U(1)$ sector.}

Solving the other half of the equations boils down to solving the unimodular constraint \eqref{eq:HTconstraint}. Again, we choose to solve this in Poincar\'{e} coordinates: using $\sqrt{g} = \tfrac{1}{z^2}$ and gauge fixing $A_z = 0$, we find that
\begin{equation}\label{eq:solutionHT}
    A_t(t,z) = \frac{1}{z} + a(t),
\end{equation}
where $a$ is a function of the Poincar\'{e} time $t$. In fact, it is rather easy to solve for $a(t)$: 
\begin{equation}
    %\overline{A}_t = f(t) \equiv \textcolor{red}{\mu} + \partial_t \varphi(t).\overline{A}_t = 
    a(t) = G^{-1}\partial_t G,
\end{equation}
where $G = e^{i\varphi}$ is a gauge transformation that maps the boundary manifold $\partial \mathcal{M}$ to group elements of $U(1)$. Note that $\varphi$ does not have to be single-valued, i.e. we have the possibility of having large gauge transformations, depending on the topology of the underlying space and whether the gauge group is compact $U(1)$ or non-compact $\mathbb{R}$. Further remarks appear in section \ref{sec:4}, where we also discuss boundary unimodular time and summarise its key properties.

Therefore, the general solution for the gauge field $A_{\mu}$ reads:
\begin{equation}
    %A_z = 0, \qquad \qquad A_t = \frac{1}{z} + \textcolor{red}{\mu}+ \partial_t \varphi(t).
    A_z = 0, \qquad \qquad A_t = \frac{1}{z} + i \,  \partial_t \varphi(t).
\end{equation}
Note that, as $z\rightarrow0$, the gauge potential becomes divergent. We can verify this by rewriting the expression in terms of the boundary parameter $u$, which yields the following asymptotic boundary condition:
\begin{equation}
    A_u = \frac{t'}{z} + i \varphi'(u) \, \, \, \, \Longrightarrow \, \, \, \,  A_u\big|_{\partial \mathcal{M}} = \frac{1}{\epsilon} + i  \varphi'(u).
\end{equation}
A final comment is that, in HT gravity, the gauge field is not pure gauge; instead, the unimodular constraint induces a boundary factor proportional to the bulk volume, scaling as $1/\epsilon$.

\paragraph{Variational principle for the HT term.} To obtain a well-posed variational principle for the bulk action \eqref{eq:bulkHT2action}, it must be supplemented with appropriate boundary contributions. In the JT sector, we fix the induced metric and the dilaton at the boundary, which necessitates the Gibbons–Hawking–York (GHY) term. Requiring a finite Euclidean action further calls for an additional counterterm. The resulting boundary terms are:
\begin{equation}
    I_{\text{bdy}} \supset - \int_{\partial \mathcal{M}} du \, \sqrt{h} \, \phi_b (K - 1),
\end{equation}
where $h$ is the induced metric determinant at the boundary. For the $U(1)$ unimodular part of the bulk action, various possible boundary conditions on the conjugate variables $\Lambda$ and $\mathcal{T}^{\mu}$ exist. Consider the variation of the bulk HT$_2$ term with respect to the conjugate pairs $\Lambda$ and $\mathcal{T}^{\mu}$, while keeping the background geometry fixed:
\begin{equation}
    \delta I_{\text{bulk}} \supset \int_{\mathcal{M}} d^2x \, \delta\Lambda \, (\sqrt{g} - \partial_{\mu} \mathcal{T}^{\mu}) - \int_{\mathcal{M}} d^2x \, \Lambda \, \partial_{\mu} \delta \mathcal{T}^{\mu}.
\end{equation}
We can perform an integration by parts on the second term, which will result in a boundary term:
\begin{align}
    \delta I_{\text{bulk}} \supset \int_{\mathcal{M}} d^2x \, \delta\Lambda \, (\sqrt{g} - \partial_{\mu} \mathcal{T}^{\mu}) + \int_{\mathcal{M}} d^2x \, (\partial_{\mu} \Lambda ) \delta\mathcal{T}^{\mu} - \int_{\partial \mathcal{M}} du\, n_{\mu}\Lambda_b \,\delta\mathcal{T}^{\mu}.
\end{align}
Here, $\Lambda_b$ is the boundary value of the vacuum cosmological constant. The on-shell action therefore reduces to having
\begin{equation}\label{eq:HTboundary}
    \delta I_{\text{bulk}}\Big|_{\text{on-shell}} \supset - \int_{\partial \mathcal{M}} du\, n_{\mu}\Lambda_b \,\delta\mathcal{T}^{\mu} = \int_{\partial \mathcal{M}} du\, \Lambda_b \, A_u.
\end{equation}
There are various ways one can eliminate the remaining boundary term to ensure a well-posed variational principle. Throughout this section, we will focus on mixed boundary conditions and analyse the implications for the boundary theory.

\subsection{Mixed boundary conditions and the complex SYK model}

We are interested in mixed boundary conditions for having a non trivial boundary dynamics for the $U(1)$ HT sector \cite{Blommaert:2018oro,Mertens:2018fds,Blommaert:2018iqz,Iliesiu:2019xuh,Iliesiu:2019lfc}. One such boundary condition is obtained by relating the boundary value of the gauge field $A_u$ to the boundary value of the vacuum cosmological constant $\Lambda_b$:
\begin{equation}\label{eq:mixedbdyconditions}
    A_u \big|_{\partial \mathcal{M}} = (- \kappa \epsilon \Lambda_b + 1)\sqrt{h}.
\end{equation}
%\FR{we say $\Lambda_b$ is a constant but we affirm this should be true $-i\varphi'=-\kappa\Lambda_b$}
This can be obtained by adding a defect term at the boundary that is quadratic in $\Lambda_b$ with coupling $\kappa$, supplemented by a counterterm that ensures the finiteness of the full boundary action:
\begin{equation}\label{eq:mixedbdyaction}
    I^{\text{mixed}}_{\text{bdy}} = - \int_{\partial \mathcal{M}} du \, \sqrt{h} \, \phi_b (K-1) - \int_{\partial \mathcal{M}} du \, \Lambda_b \,\bigl(A_u  - \sqrt{h}\bigr) - \frac{\kappa \epsilon}{2} \int_{\partial \mathcal{M}} du \, \sqrt{h} \, \Lambda_b^2.
\end{equation}In fact, the variation of the bulk HT$_2$ action gives a boundary variation proportional to $\delta \mathcal{T}^{\mu}$, which is cancelled by a variation of $I^{\text{mixed}}_{\text{bdy}}$ in \eqref{eq:HTboundary}. The remaining variation with respect to the vacuum cosmological constant $\delta \Lambda_b$ imposes the mixed boundary conditions given above to vanish. Hence, including this boundary term in the action renders the variational principle well-posed. The choice of mixed boundary condition \eqref{eq:mixedbdyconditions} is not arbitrary: in section \ref{sec:4}, we will provide further physical context for it.

It is rather useful to complete the square in $\Lambda_b$ for this boundary integral. The resulting action is
\begin{align}
    I^{\text{mixed}}_{\text{bdy}} &= -\frac{\kappa \epsilon}{2} \int_{\partial \mathcal{M}} du \, \sqrt{h}\, \left(\Lambda_b
    + \frac{A_u - \sqrt{h}}{\kappa \epsilon \sqrt{h}}\right)^2 
    + \frac{1}{2 \kappa \epsilon} \int_{\partial \mathcal{M}} du \, \sqrt{h_{uu}}h^{uu} (A_u - \sqrt{h})^2 \nonumber\\
    & - \int_{\partial \mathcal{M}} du \, \sqrt{h} \, \phi_b (K-1),
\end{align}
where we used $\tfrac{1}{\sqrt{h}} = \sqrt{h_{uu}} \, h^{uu}$. Plugging in the boundary condition \eqref{eq:mixedbdyconditions} relating $A_u|_{\partial \mathcal{M}}$ and $\Lambda_b$ results in the on-shell action to be simply
\begin{equation}\label{eq:mixedonshellaction}
    I^{\text{mixed}}\Big|_{\text{on-shell}} =  - \int_{\partial \mathcal{M}} du \, \sqrt{h} \, \phi_b (K-1) + \frac{1}{2 \kappa \epsilon} \int_{\partial \mathcal{M}} du \, \sqrt{h_{uu}} h^{uu} \, (A_u - \sqrt{h})^2.
\end{equation}
The same result follows by imposing the mixed boundary conditions \eqref{eq:mixedbdyconditions} directly in the boundary action \eqref{eq:mixedbdyaction}.

With the correct boundary terms identified, we can evaluate the on-shell action. The first term in this boundary action in Poincar\'{e} coordinates gives the well-known Schwarzian \cite{Almheiri:2014cka,Maldacena:2016hyu,Jensen:2016pah,Maldacena:2016upp,Engelsoy:2016xyb,Saad:2019lba,Stanford:2019vob,Mertens:2022irh}\footnote{The Schwarzian derivative for a function $F(x)$ is defined as follows:
\begin{equation}
    \text{Sch}\{F,x\} := \left(\frac{F''}{F'} \right)' - \frac{1}{2} \left(\frac{F''}{F'}\right)^2 \equiv \frac{F'''}{F'} -\frac{3}{2} \left(\frac{F''}{F'}\right)^2.
\end{equation}
Crucially, it is invariant under M\"{o}bius (or $SL(2,\mathbb{R})$) transformations $F \rightarrow \tfrac{aF + b}{cF + d}$ with $ad - bc = 1$.}:
\begin{equation}
    K - 1 = \epsilon^2 \,\text{Sch}\{ t(u),u\} + \mathcal{O}(\epsilon^4).
\end{equation}
Then, by considering the asymptotic solution for $A_u$ at the boundary:
\begin{equation}\label{eq:gaugefieldasymptoticsolutionPoincare}
    A_u\big|_{\partial \mathcal{M}} = \frac{1}{\epsilon} + i \varphi'(u),
\end{equation}
we find that the on-shell action becomes:
\begin{equation}\label{eq:HT2zerotempcSYK}
    I^{\text{mixed}}\Big|_{\text{on-shell}} =  - \overline{\phi}_r \int_{\partial \mathcal{M}} du \, \, \text{Sch}\{t,u\} - \frac{1}{2 \kappa} \int_{\partial \mathcal{M}} du \, \varphi'^2.
\end{equation}

As expected, the first term reproduces the Schwarzian action, while the second describes a “massless” free particle on the group manifold $U(1)$; remarkably, this boundary theory has been studied in the literature and identified with the low-energy effective action of the complex SYK model \cite{Davison:2016ngz,Mertens:2019tcm}. This is not merely coincidental, and in section \ref{sec:3}, we elucidate its origin by exploring the relation between 2d Maxwell theory and HT$_2$ gravity. In fact, as we show in section \ref{sec:4}, adding the defect term promotes the boundary dynamics to a $(0+1)$-dimensional Henneaux–Teitelboim unimodular theory coupled to the Schwarzian mode, in which the mixed boundary condition is naturally associated with the boundary unimodular HT constraint.

\paragraph{Comment on counterterms.} 

We would like to add comments regarding the choice of mixed boundary conditions. We start off by rewriting the boundary condition as
\begin{equation}
    A_u \big|_{\partial \mathcal{M}} - \frac{1}{\epsilon}= - \kappa \Lambda_b.
\end{equation}
Then, one may see that the left hand side is nothing but the renormalised gauge potential $\overline{A}_u$ evaluated at the boundary. These mixed boundary conditions can be derived directly from the following boundary action:
\begin{equation}
    I^{\text{mixed}}_{\text{bdy}} = - \int_{\partial \mathcal{M}} du \, \sqrt{h} \, \phi_b (K-1) - \int_{\partial \mathcal{M}} du \, \Lambda_b \overline{A}_u - \frac{\kappa \epsilon}{2} \int_{\partial \mathcal{M}} du \, \sqrt{h} \, \Lambda_b^2.
\end{equation}
Indeed, the variation of this action with respect to $\Lambda_b$ gives the mixed boundary conditions for the renormalised gauge potential. The counterterm being absorbed directly in the renormalised gauge field, the resulting on-shell action is therefore identical to that of the Schwarzian coupled to the action of a free particle on the $U(1)$ manifold.

\subsection{Thermal aspects of HT\texorpdfstring{$_2$}{HT2} gravity}

To study finite-temperature systems, Poincaré coordinates are unsuitable, since their horizon is at zero temperature. A non-zero temperature is obtained by introducing Rindler (thermal) coordinates $(\tau,\rho)$, defined through the transformations:
\begin{equation}\label{eq:PoincaretoRindler}
    z \mp i t = \frac{e^{-i\frac{\pi \tau}{\beta}}(1 \pm \cosh \rho) - e^{i\frac{\pi \tau}{\beta}}\sinh \rho}{e^{-i\frac{\pi \tau}{\beta}}(1 \pm \cosh \rho) + e^{i\frac{\pi \tau}{\beta}}\sinh \rho}.    
\end{equation}
The resulting Euclidean metric takes the form \cite{Mertens:2022irh}:
\begin{equation}
    ds^2 = r_h^2\sinh^2(\rho) \, d\tau^2 + d\rho^2,
\end{equation}
with $\rho$ as the proper distance coordinate to the black hole horizon $r_h = \tfrac{2 \pi}{\beta}$. In these coordinates, the asymptotic boundary is located at $\rho \rightarrow \infty$. The Euclidean metric covers the entire hyperbolic disk, with periodic Rindler time $\tau \sim \tau + \beta$. The boundary manifold is at a finite temperature $\beta$, and denoted by $\partial \mathcal{M} \cong S^1_\beta$. Alternatively, one can also consider the static black hole coordinates $(\tau,r)$ with radial gauge $\phi = r$ for the dilaton and line element given by:
\begin{equation}\label{eq:bhmetric}
    ds^2 = (r^2 - r_h^2) \,d\tau^2 + \frac{dr^2}{r^2 - r_h^2},
\end{equation}
such that the Hawking temperature is $T_H = \beta^{-1} = \tfrac{r_h}{2\pi}$.\\

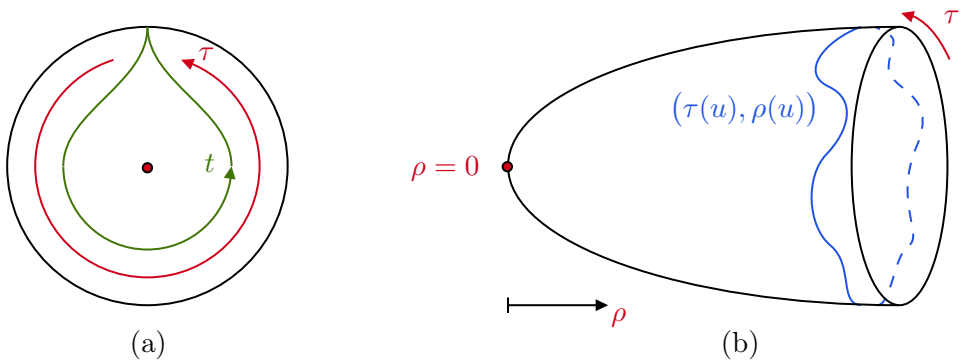
\begin{figure}[!ht]
\centering

\tikzset{every picture/.style={line width=0.75pt}} %set default line width to 0.75pt        

\begin{tikzpicture}[x=0.75pt,y=0.75pt,yscale=-1,xscale=1]
%uncomment if require: \path (0,208); %set diagram left start at 0, and has height of 208

%Shape: Ellipse [id:dp9577334201890827] 
\draw   (110,80) .. controls (110,41.34) and (141.34,10) .. (180,10) .. controls (218.66,10) and (250,41.34) .. (250,80) .. controls (250,118.66) and (218.66,150) .. (180,150) .. controls (141.34,150) and (110,118.66) .. (110,80) -- cycle ;
%Shape: Arc [id:dp8233892464354852] 
\draw  [draw opacity=0] (196.79,26.56) .. controls (219.51,33.69) and (236,54.92) .. (236,80) .. controls (236,110.93) and (210.93,136) .. (180,136) .. controls (149.07,136) and (124,110.93) .. (124,80) .. controls (124,54.98) and (140.41,33.79) .. (163.06,26.61) -- (180,80) -- cycle ; \draw [color={rgb, 255:red, 208; green, 2; blue, 27 }  ,draw opacity=1 ]   (199.82,27.61) .. controls (220.96,35.61) and (236,56.05) .. (236,80) .. controls (236,110.93) and (210.93,136) .. (180,136) .. controls (149.07,136) and (124,110.93) .. (124,80) .. controls (124,54.98) and (140.41,33.79) .. (163.06,26.61) ;  \draw [shift={(196.79,26.56)}, rotate = 21.83] [fill={rgb, 255:red, 208; green, 2; blue, 27 }  ,fill opacity=1 ][line width=0.08]  [draw opacity=0] (6.25,-3) -- (0,0) -- (6.25,3) -- cycle    ;
%Shape: Ellipse [id:dp27511100770186425] 
\draw  [fill={rgb, 255:red, 208; green, 2; blue, 27 }  ,fill opacity=1 ] (177.67,80.88) .. controls (177.67,79.59) and (178.71,78.54) .. (180,78.54) .. controls (181.29,78.54) and (182.33,79.59) .. (182.33,80.88) .. controls (182.33,82.16) and (181.29,83.21) .. (180,83.21) .. controls (178.71,83.21) and (177.67,82.16) .. (177.67,80.88) -- cycle ;
%Shape: Arc [id:dp36751114996814915] 
\draw  [draw opacity=0] (222,80.15) .. controls (221.92,103.28) and (203.15,122) .. (180,122) .. controls (156.88,122) and (138.13,103.33) .. (138,80.24) -- (180,80) -- cycle ; \draw [color={rgb, 255:red, 65; green, 117; blue, 5 }  ,draw opacity=1 ]   (221.88,83.24) .. controls (220.22,104.92) and (202.11,122) .. (180,122) .. controls (156.88,122) and (138.13,103.33) .. (138,80.24) ;  \draw [shift={(222,80.15)}, rotate = 98.52] [fill={rgb, 255:red, 65; green, 117; blue, 5 }  ,fill opacity=1 ][line width=0.08]  [draw opacity=0] (6.25,-3) -- (0,0) -- (6.25,3) -- cycle    ;
%Curve Lines [id:da8525906535186417] 
\draw [color={rgb, 255:red, 65; green, 117; blue, 5 }  ,draw opacity=1 ]   (138,80) .. controls (138.2,52.42) and (180.2,38.42) .. (180,10) ;
%Curve Lines [id:da7208048885278961] 
\draw [color={rgb, 255:red, 65; green, 117; blue, 5 }  ,draw opacity=1 ]   (222,80) .. controls (221.8,51.62) and (180.6,38.42) .. (180,10) ;
%Curve Lines [id:da5815737633260927] 
\draw [color={rgb, 255:red, 30; green, 83; blue, 227 }  ,draw opacity=1 ][line width=0.75]    (536,150) .. controls (525,149.09) and (532.67,129.76) .. (521,120) .. controls (509.33,110.24) and (508.2,85.22) .. (519,74) .. controls (529.8,62.78) and (534.6,46.82) .. (520,40) .. controls (505.4,33.18) and (517.67,17.09) .. (536,10) ;
%Curve Lines [id:da36434829989773476] 
\draw [color={rgb, 255:red, 30; green, 83; blue, 227 }  ,draw opacity=1 ][line width=0.75]  [dash pattern={on 4.5pt off 4.5pt}]  (544,150) .. controls (551.67,150.42) and (552.33,125.09) .. (560,120) .. controls (567.67,114.91) and (552.33,97.76) .. (564,70) .. controls (575.67,42.24) and (549.4,42.42) .. (550,30) .. controls (550.6,17.58) and (558.2,16.42) .. (543,10) ;
%Shape: Arc [id:dp9931794553355715] 
\draw  [draw opacity=0] (556.23,3.05) .. controls (565.74,3.75) and (574.28,12.58) .. (580.4,26.39) -- (555,78) -- cycle ; \draw [color={rgb, 255:red, 208; green, 2; blue, 27 }  ,draw opacity=1 ]   (559.19,3.53) .. controls (567.5,5.66) and (574.92,14.03) .. (580.4,26.39) ;  \draw [shift={(556.23,3.05)}, rotate = 17.58] [fill={rgb, 255:red, 208; green, 2; blue, 27 }  ,fill opacity=1 ][line width=0.08]  [draw opacity=0] (6.25,-3) -- (0,0) -- (6.25,3) -- cycle    ;
%Shape: Arc [id:dp6521489033942849] 
\draw  [draw opacity=0] (555.07,150) .. controls (555.05,150) and (555.02,150) .. (555,150) .. controls (447.3,150) and (360,118.66) .. (360,80) .. controls (360,41.34) and (447.3,10) .. (555,10) .. controls (555.15,10) and (555.29,10) .. (555.44,10) -- (555,80) -- cycle ; \draw   (555.07,150) .. controls (555.05,150) and (555.02,150) .. (555,150) .. controls (447.3,150) and (360,118.66) .. (360,80) .. controls (360,41.34) and (447.3,10) .. (555,10) .. controls (555.15,10) and (555.29,10) .. (555.44,10) ;  
%Shape: Ellipse [id:dp23772728340754978] 
\draw   (555.44,10) .. controls (568.64,10) and (579.35,41.34) .. (579.35,80) .. controls (579.35,118.66) and (568.64,150) .. (555.44,150) .. controls (542.24,150) and (531.53,118.66) .. (531.53,80) .. controls (531.53,41.34) and (542.24,10) .. (555.44,10) -- cycle ;
%Shape: Ellipse [id:dp5550098592658707] 
\draw  [fill={rgb, 255:red, 208; green, 2; blue, 27 }  ,fill opacity=1 ] (357.33,80.33) .. controls (357.33,79.04) and (358.38,78) .. (359.67,78) .. controls (360.96,78) and (362,79.04) .. (362,80.33) .. controls (362,81.62) and (360.96,82.67) .. (359.67,82.67) .. controls (358.38,82.67) and (357.33,81.62) .. (357.33,80.33) -- cycle ;
%Straight Lines [id:da7012758003886045] 
\draw    (360,150) -- (407,150) ;
\draw [shift={(410,150)}, rotate = 180] [fill={rgb, 255:red, 0; green, 0; blue, 0 }  ][line width=0.08]  [draw opacity=0] (6.25,-3) -- (0,0) -- (6.25,3) -- cycle    ;
\draw [shift={(360,150)}, rotate = 180] [color={rgb, 255:red, 0; green, 0; blue, 0 }  ][line width=0.75]    (0,3.91) -- (0,-3.91)   ;

% Text Node
\draw (207,72.4) node [anchor=north west][inner sep=0.75pt]  [color={rgb, 255:red, 65; green, 117; blue, 5 }  ,opacity=1 ]  {$t$};
% Text Node
\draw (204,18.7) node [anchor=north west][inner sep=0.75pt]  [color={rgb, 255:red, 208; green, 2; blue, 27 }  ,opacity=1 ]  {$\tau $};
% Text Node
\draw (478,51.5) node  [color={rgb, 255:red, 30; green, 83; blue, 227 }  ,opacity=1 ]  {$\bigl( \tau (u) ,\rho ( u)\bigr)$};
% Text Node
\draw (180.5,170) node   [align=left] {(a)};
% Text Node
\draw (476,170) node   [align=left] {(b)};
% Text Node
\draw (310,73.4) node [anchor=north west][inner sep=0.75pt]    {$\textcolor[rgb]{0.82,0.01,0.11}{\rho =0}$};
% Text Node
\draw (576,0.4) node [anchor=north west][inner sep=0.75pt]  [color={rgb, 255:red, 208; green, 2; blue, 27 }  ,opacity=1 ]  {$\tau $};
% Text Node
\draw (410,150.4) node [anchor=north west][inner sep=0.75pt]    {$\textcolor[rgb]{0.82,0.01,0.11}{\rho }$};

\end{tikzpicture}
    \caption{Panel (a) shows the coordinate systems on the hyperbolic disk. Panel (b) shows the standard deformation to the cigar geometry; the Schwarzian mode lives on the cutoff boundary, shown as the blue circle.}
    %\label{fig:placeholder}
\end{figure}
In the previous section, the Schwarzian action (and thus the effective complex SYK action) was written at zero temperature. To obtain its finite-temperature counterpart, we use the relation between boundary Poincaré and Rindler times. Taking the boundary limit of \eqref{eq:PoincaretoRindler} by sending $\rho \to \infty$ and $z \to 0$, the two time coordinates are related by
\begin{equation}
    t(u) = \tan \left(\frac{\pi}{\beta} f(u)\right).
\end{equation}
where we used the time-reparametrization $\tau \mapsto f(u)$ of the boundary thermal circle. The periodicity condition $f(u+\beta) = f(u) + \beta$ and $f'\geq 0$ ensures that $f \in \mathrm{Diff}(S^1_{\beta})$. It is now easy to see that, the Schwarzian at finite-temperature takes the form:
\begin{align}\label{eq:thermalSchwarzian}
    I_{\text{JT}}\Big|_{\text{on-shell}} &=- \overline{\phi}_r \int_0^\beta du \, \,  \text{Sch}\left\{\tan\left(\frac{r_h}{2} f(u) \right),u\right\} \\
    &=- \overline{\phi}_r \int_0^\beta du \, \, \left(\frac{r_h^2}{2} f'^2+ \text{Sch}\bigl\{f,u\bigr\} \right).
\end{align}

In what follows, we derive the finite-temperature analogue of the action \eqref{eq:HT2zerotempcSYK} from the boundary dynamics of HT$_2$ gravity, working in black hole coordinates. Notice that in these thermal coordinates, the metric determinant is simply gauge fixed to be $\sqrt{g} = 1$. In these coordinates, the HT constraint reduces to
\begin{equation}
    \varepsilon^{\mu \nu} \partial_\mu A_\nu = 1.
\end{equation}
A gauge field configuration that satisfies this equation is given by
\begin{equation}
    A_{r}(\tau,r) = i \partial_{r} \Phi(\tau,r) ,\,\,\,\,\,\,\,\,\, A_{\tau}(\tau,r) = r - r_h + i \partial_{\tau} \Phi(\tau,r),
\end{equation}
for some function $\Phi(\tau,r)$. The solution for $A_\tau$ satisfies $A_\tau(\tau,r_h)=0$, which guarantees smoothness at the horizon $r=r_h$ and makes the potential difference in $A_\tau$ a gauge-invariant quantity \cite{Sachdev:2019bjn}.

Evaluating the gauge field on the thermal circle, we find that the asymptotic boundary condition for the gauge potential is
\begin{equation}\label{eq:thermalgaugefieldsolution}
    A_u\big|_{S^1_{\beta}} = \frac{1}{\epsilon} - r_hf' + i \varphi'(u),
\end{equation}
with the condition that $\Phi(u,r_{\infty}) \equiv \varphi(u)$ at the boundary, such that the phase field defines a boundary Wilson line extending from $r_h$ to the boundary \cite{Sachdev:2019bjn}\footnote{In these coordinates, it is simple to check that the phase field satisfies the periodicity condition
\begin{equation}
    \Delta \varphi := \int_0^{\beta} du \, \varphi'(u) = \varphi(\beta) - \varphi(0) \in 2 \pi \mathbb{Z}.
\end{equation}}:
\begin{equation}
    \varphi(u) \equiv \Phi(u,r_{\infty}) := -i \int_{r_h}^{r_{\infty}} dr \, A_r(u,r).
\end{equation}

We can obtain the thermal effective action in close analogy with the zero-temperature on-shell action \eqref{eq:HT2zerotempcSYK}. Starting from the mixed boundary condition on-shell action \eqref{eq:mixedonshellaction}, we substitute the solution for the extrinsic curvature with the thermal identification $t(u) = \tan\frac{\pi}{\beta} f(u)$, which produces the thermal Schwarzian term in \eqref{eq:thermalSchwarzian}. We then insert the asymptotic thermal gauge field configuration given by \eqref{eq:thermalgaugefieldsolution}. The resulting on-shell action is the effective action of thermal complex SYK:
\begin{equation}\label{eq:mixedthermalcSYK}
    I^{\text{mixed}}\Big|_{\text{on-shell}} = - \overline{\phi}_r \int_0^\beta du \, \, \left(\frac{r_h^2}{2} f'^2+ \text{Sch}\bigl\{f,u\bigr\} \right) - \frac{1}{2 \kappa} \int_0^\beta du \, \bigl(\varphi' + i r_h f'\bigr)^2.
\end{equation}
This action enjoys a global $SL(2,\mathbb{R}) \times U(1)$ symmetry, and the theory may be viewed as describing the pseudo-Goldstone dynamics associated with the breaking of time reparametrizations, together with time-dependent phase rotations on the thermal circle \cite{Godet:2020xpk}. The symmetry-breaking pattern\footnote{An equivalent viewpoint comes from the structure of asymptotic symmetries and their algebras, for which we refer the reader to \cite{Godet:2020xpk}.} is
\begin{equation}
    \mathrm{Diff}(S^1_{\beta}) \ltimes LU(1) \longrightarrow SL(2,\mathbb{R}) \times U(1).
\end{equation}
It is easy to see that this action can also be obtained from the zero-temperature on-shell action \eqref{eq:HT2zerotempcSYK} by performing the shift in the phase field $\varphi \rightarrow \varphi + i r_hf'$, which amounts to coupling the boundary time reparametrization mode $f$ to the phase field $\varphi$. It is also noteworthy that in the zero-temperature limit, \eqref{eq:mixedthermalcSYK} reduces to the zero-temperature result \eqref{eq:HT2zerotempcSYK}, where one takes $\beta \rightarrow \infty$ (or $r_h \rightarrow 0$).

\subsection{Thermoelectric properties of the boundary theory}\label{sec:2.3}

In the complex SYK model, the fermion number defines the charge $\mathcal{Q}$, which in turn appears as the thermodynamic conjugate to the chemical potential $\mu$. From Maxwell’s thermodynamic relations, the change in entropy per unit charge $\mathcal{Q}$ is tied to the slope of the chemical potential versus temperature at fixed charge \cite{Sachdev:2019bjn}:
\begin{equation}
    2 \pi \mathcal{E} = \frac{dS}{d\mathcal{Q}} = - \lim_{T \rightarrow 0} \left(\frac{\partial \mu}{\partial T}\right)_{\mathcal{Q}},
\end{equation}
with $\mathcal{E}$ the dimensionless electric field in the $AdS_2$ region (also known as the particle-hole asymmetry). Indeed, in the $T \rightarrow0$ limit, the entropy as a function of fermion number $\mathcal{Q}$ is
\begin{equation}\label{eq:entropy}
    S(\mathcal{Q},T\rightarrow 0) = S_0 + 2 \pi \mathcal{E} \mathcal{Q},
\end{equation}
where $S_0$ is the value of entropy at extremality. On the other hand, at fixed charge $\mathcal{Q}$ and in the limit $T \rightarrow0$, the chemical potential takes the form
\begin{equation}
    \mu(\mathcal{Q},T \rightarrow0) = \mu_0 - 2 \pi \mathcal{E} T,
\end{equation}
where $\mu_0$ is the chemical potential at extremality and zero-temperature. Since the Schwarzian-like part of the effective action only captures deviations from extremality, we shall not be worried about the extremal quantities $S_0$ and $\mu_0$. 

We can then match the dimensionful parameters of the complex SYK model—such as the heat capacity $\gamma$, compressibility $K$, chemical potential $\mu$, and inverse temperature $\beta$—to the corresponding quantities appearing in the boundary on-shell action of HT$_2$ gravity. This suggests that we have:
\begin{equation}
    \overline{\phi}_r = \frac{\gamma}{4 \pi^2}, \ \ \ \ \ \ \ \ \frac{1}{\kappa} = K, \ \ \ \ \ \ \ \ r_h = -\mu \,\,\, (\text{with} \ \mathcal{E} = 1), \ \ \ \ \ \ \ \beta \equiv 1/T = \frac{2\pi}{r_h}.
\end{equation}
In this regard, the electric field---or the chemical potential---in the effective complex SYK action controls the coupling between energy and fermion number fluctuations, described by $f(u)$ and $\varphi(u)$, respectively. For the HT$_2$ boundary theory, this fixes the electric field to the specific value $\mathcal{E}=1$ (or $\mu = -2\pi T$). 

%Likewise, away from extremality, the entropy \eqref{eq:entropy} becomes directly proportional to $\mathcal{Q}$. 

The identification $\mu = -r_h$ follows directly from the HT constraint. Written in terms of differential forms, we have
\begin{equation}
    F = \hodge \mathbf{1} \qquad \Longleftrightarrow \qquad \hodge F = 1.
\end{equation}
This allows us to identify a constant electric field such that $\mathcal{E} = 1$. In fact, the HT constraint equation also satisfies a Maxwell-like equation $d \, \hodge F = 0$; hence restricting the electric field. Using the general relation $\mu = - 2\pi \mathcal{E} T$ for the chemical potential in complex SYK model, we recover precisely the expression appearing in \eqref{eq:mixedthermalcSYK}. 

Another way to see the particular value of the electric field is to consider the $AdS_2$ near-horizon region of asymptotically $AdS_4$ black holes at near extremality \cite{Davison:2016ngz,Sachdev:2019bjn}. Then, we can write the gauge field $\widetilde{A}_\tau$ solving the 2d Maxwell-dilaton-gravity theory given by
\begin{equation}
    \widetilde{A}_\tau = \mathcal{E} (r - r_h),
\end{equation}
and with the pure gauge configuration $\widetilde{A}_r = i \partial_r \Phi$. Note that this is precisely the HT$_2$ gauge field solution \eqref{eq:thermalgaugefieldsolution} with $\widetilde{A}_\tau = \mathcal{E}(A_\tau - i \partial_\tau \Phi)$. The fact that we find a particular value for the electric field and chemical potential on the boundary theory will be further explored in the next section.

\section{HT\texorpdfstring{$_2$}{HT2} gravity, 2d Maxwell and BF gauge theory}\label{sec:3}

In the preceding sections, we saw that HT$_2$ gravity naturally leads to a boundary theory captured by the low-energy effective action of the complex SYK model. In the literature, this structure is usually discussed in the setting of JT gravity coupled to 2d Maxwell theory. This agreement is not accidental: two-dimensional gravity is highly constrained, and the Schwarzian mode (together with its $U(1)$ extension in the presence of a conserved current) is universal for near-CFT$_1$ systems with the appropriate symmetries. This section is not about claiming this surprising coincidence; rather, it precisely clarifies how seemingly distinct bulk formulations realise the same universal boundary dynamics: the Schwarzian mode together with a free particle on $U(1)$. 

We begin by showing that 2d Maxwell theory can be recast in HT form, motivated by the four-dimensional literature of Hawking and Kaloper that makes use of top forms. We then turn to HT$_2$ gravity and explore it as a BF gauge theory, making the underlying algebraic structure manifest. Finally, in the last subsection, these results enable us to discuss how HT$_2$ gravity can be embedded within higher-dimensional black hole physics.

The central point of this section is to highlight a connection that appears to be largely absent from the existing discussions: 2d Maxwell theory is a top-form theory, and therefore, encodes the vacuum energy. In fact, this closely mirrors the role played by top form actions in four-dimensional gravity, as first raised by Hawking (see, for instance, \cite{Aurilia:1980xj,Hawking:1984hk,Duff:1989ah}). We will adopt this perspective, as it provides a natural intuition for the vacuum contribution to the cosmological constant in two dimensions. The theory of relevance is JT gravity coupled to Maxwell theory, with action given by:
\begin{align}
    I_{\text{JT+Maxwell}} = -\frac{1}{2} \int_{\mathcal{M}} d^2x \, \sqrt{g}\, \phi (R+2) + \frac{1}{4e^2}\int_{\mathcal{M}} d^2x \, \sqrt{g} \, f_{\mu \nu}f^{\mu \nu} + I_{\text{bdy}}.
\end{align}
The field strength is a 2-form, and on-shell it is locally proportional to the volume form. As a result, its contribution acts as an effective vacuum energy and hence directly shifts the cosmological constant. We will make contact with the four-dimensional analogue, where the cosmological constant can likewise be associated with a top form field strength.

It is known that in four dimensions, the vacuum cosmological constant can be recast in terms of top forms \cite{Aurilia:1980xj,Hawking:1984hk,Duff:1989ah}. Consider the $4d$ Einstein-Hilbert action with a Maxwell-like term:
\begin{equation}\label{eq:hawking}
    I_{\text{H}} = - \frac{1}{2} \int_{\mathcal{M}} d^4x \, \sqrt{g}\, R + \frac{1}{48} \int_{\mathcal{M}} d^4x \, \sqrt{g} \, f_{\mu \nu \lambda \sigma} f^{\mu \nu \lambda \sigma},
\end{equation}
where $f_{\mu \nu \lambda \sigma} = 4 \partial_{[\mu} a_{\nu \lambda \sigma]}$ is a top form field strength for the 3-form gauge field $a_{\mu \nu \lambda}$. Variation with respect to the gauge field gives the Maxwell-like equation
\begin{equation}
    \nabla_{\mu} f^{\mu \nu \lambda \sigma} = 0.
\end{equation}
In four dimensions, a 4-form is necessarily proportional to the volume form, $f_{\mu \nu \lambda \sigma}= f(x)\,\epsilon_{\mu \nu \lambda \sigma}$, so the equation above implies that $f(x)=f_0$ is a constant. Substituting this solution into the action yields an effective cosmological constant $\Lambda_{\text{eff}}=\tfrac{f_0^2}{2}$. 
\begin{equation}
    I_{\text{H}} \Big|_{\text{on-shell}} = - \frac{1}{2} \int_{\mathcal{M}} d^4x \, \sqrt{g}\, \left(R - 2  \Lambda_{\text{eff}}\right).
\end{equation}
However, the effective cosmological constant acquires the correct sign only once the action is supplemented with the appropriate boundary terms, as emphasised by Duff in \cite{Duff:1989ah}. While this point is central to their analysis, our focus here is instead on the structural link between the cosmological constant and top form gauge fields.

Recently, it has been shown by Kaloper \cite{Kaloper:2023xfl} that the Maxwell-like term can be recast as an HT unimodular term. Starting from \eqref{eq:hawking}, we can introduce a scalar field $\chi$ such that the action takes the form
\begin{equation}\label{eq:quadraticKaloperaction}
    I_{\text{K}} = -\frac{1}{2} \int_{\mathcal{M}} d^4x \, \sqrt{g} \, R - 2 \int_{\mathcal{M}} d^4x \, \left( \chi \, \varepsilon^{\mu \nu \lambda \sigma} f_{\mu \nu \lambda \sigma} + \chi^2\sqrt{g}\right) + I_{\text{bdy}}.
\end{equation}
This action is equivalent to the Maxwell-like term upon Gaussian integration with respect to $\chi$. Using the following field redefinition of the fields $(\chi,a_{\mu \nu \lambda})$:
\begin{equation}
    \Lambda = -2 \chi^2, \, \, \, \, \, \, \, \, \, \, \, \, \, \, \, \, \, A_{\mu \nu \lambda} = - \frac{12 \, a_{\mu \nu \lambda}}{\chi},
\end{equation}
the resulting bulk action becomes HT gravity in four dimensions, that is:
\begin{equation}
    I_{\text{K}} -\frac{1}{2} \int_{\mathcal{M}} d^4x \, \sqrt{g} \, R - \int_{\mathcal{M}} d^4x \, \Lambda \left(\frac{1}{24} \varepsilon^{\mu \nu \lambda \sigma} F_{\mu \nu \lambda \sigma} - \sqrt{g}\right) + I_{\text{bdy}},
\end{equation}
where $F_{\mu \nu \lambda \sigma} = 4 \partial_{[\mu} A_{\nu \lambda \sigma]}$ and $\mathcal{T}^{\mu} = \tfrac{1}{3!}\varepsilon^{\mu \nu \lambda \sigma} A_{\nu \lambda \sigma}$ can be seen as the densitized dual of $A_{\mu \nu \lambda}$ that appears in the standard HT gravity action \eqref{eq:HTgravityaction}.

Hence we can use this to recast 2d Maxwell action as an HT term, and we provide an alternative perspective on why the two theories share similar holographic descriptions via the effective action for complex SYK model. Throughout this section we will also mention when relevant the relation with flat HT$_2$ (or $\reallywidehat{\text{CGHS}}$ model).

%Further more we will use these techniques to find 0+1 HT at the boundary in section 4. 

\subsection{Bridging 2d Maxwell to HT\texorpdfstring{$_2$}{HT2}}

Here, we make use of Kaloper's field redefinition explained previously within the two-dimensional context, and further give illuminating insights about the boundary theory in regards of HT$_2$ and its connection with the particle on $U(1)$.

Consider JT gravity coupled to 2d Maxwell theory:
\begin{equation}\label{eq:JT2dMaxwell}
    I = -\frac{1}{2} \int_{\mathcal{M}} d^2x \, \sqrt{g} \, \phi (R+2) + \frac{1}{4 e^2} \int_{\mathcal{M}} d^2x \, \sqrt{g} \, f_{\mu \nu} f^{\mu \nu},
\end{equation}
where $f=da$ is the Maxwell field strength of the $U(1)$ gauge field $a$, and $e$ is the coupling constant of Maxwell theory. We have omitted boundary terns for the time being, and will come back to this shortly. Since 2d Maxwell field strength is a top form on $\mathcal{M}$, we can always write it as an antisymmetric tensor proportional to the Levi-Civita tensor $f_{\mu \nu} = \mathcal{E}(x) \epsilon_{\mu\nu}$. Then, from Maxwell's equations:
\begin{equation}
    \nabla_{\mu} f^{\mu \nu} = 0,
\end{equation}
we find that the electric field $\mathcal{E}(x) = \mathcal{E}$ is a constant of motion. This is akin to the Hawking four-dimensional construction of an effective vacuum cosmological constant from a 4-form field strength discussed previously. Indeed, we find that on-shell, we get an effective vacuum cosmological constant with $\Lambda_{\text{eff}} = \tfrac{\mathcal{E}^2}{2e^2}$.
 
One may instead formulate 2d Maxwell theory in terms of$\; $``dual variables'', leading to the equivalent action \cite{Witten:1991we,Blommaert:2018oue}:
\begin{equation}\label{eq:quadraticaction}
    I \supset - \int_{\mathcal{M}} d^2x \, \chi \, \varepsilon^{\mu \nu} f_{\mu \nu} - 2e^2 \int_{\mathcal{M}} d^2x \, \sqrt{g} \, \chi^2,
\end{equation}
where we omit the JT gravity part of the action and focus solely on the gauge theory sector. The above action is equivalent to 2d Maxwell theory upon performing the Gaussian integration on $\chi$. Variation with $\chi$ gives:
\begin{equation}\label{eq:chiequation}
    \chi = - \frac{1}{4 e^2} \epsilon^{\mu \nu} f_{\mu\nu},
\end{equation}
and the variation with respect to $a_{\mu}$ is that $\chi$ is a constant of motion $\chi = \chi_0$. We once again find an effective vacuum cosmological constant given by
\begin{equation}
     \Lambda_{\text{eff}} = 2e^2 \chi_0^2 \equiv \frac{\mathcal{E}^2}{2 e^2}.
\end{equation}

We now want to show how the quadratic $\chi^2$-action is related to HT$_2$ gravity. For this, we consider the following field redefinition, akin to Kaloper \cite{Kaloper:2023xfl}:
\begin{equation}\label{eq:fieldredefinition}
    \Lambda = -2 e^2 \chi^2, \, \, \, \, \, \, \, \, \, \, \, \, \, \,  A_{\mu} = - \frac{a_{\mu}}{2 e^2 \chi}.
\end{equation}
Up to a possible boundary term, we remarkably find that the quadratic action for the dual variable formulation of 2d Maxwell theory reduces to a linear action in $\Lambda$, and gives exactly the bulk HT$_2$ gravity action, namely:
\begin{equation}
    I = -\frac{1}{2} \int_{\mathcal{M}} d^2x \, \sqrt{g} \, \phi (R+2) - \int_{\mathcal{M}} d^2x \, \Lambda \, \varepsilon^{\mu \nu} \partial_{\mu} A_{\nu} + \int_{\mathcal{M}} d^2x \, \sqrt{g} \, \Lambda \equiv I_{\text{HT}_2}.
\end{equation}
Interestingly, the same field redefinition was already observed in the flat JT limit of the $\widehat{\mathrm{CGHS}}$ framework \cite{Godet:2021cdl}, although it was not recognized there as a general feature of flat HT$_2$. For our purposes, it is necessary to extend the analysis to non-flat backgrounds, where boundary contributions play a crucial role in relating 2d Maxwell theory to HT$_2$.

\paragraph{Boundary consideration.}

It would be useful to check how HT$_2$ action ties with the Maxwell theory and its boundary description as the effective action for the universal sector of complex SYK. To this end, consider the full HT$_2$ gravity action with appropriate boundary terms and boundary defect given in \eqref{eq:mixedbdyaction}:
\begin{align}\label{eq:lambdaaction}
    I \supset  \int_{\mathcal{M}} d^2x \, \Lambda (\sqrt{g} - \varepsilon^{\mu \nu} \partial_{\mu}A_{\nu}) - \int_{\partial \mathcal{M}} du \, \Lambda_b (A_u-\sqrt{h}) - \frac{\kappa \epsilon}{2} \int_{\partial \mathcal{M}} du \, \sqrt{h} \, \Lambda_b^2.
\end{align}
Now, by performing the field redefinitions given by \eqref{eq:fieldredefinition}, we find that after integration by parts and correctly manipulating the boundary terms, the resulting action is:
\begin{equation}
    I \supset 2 \int_{\mathcal{M}} d^2x \, \bigl( \varepsilon^{\mu \nu} (\partial_{\mu} \chi) a_{\nu} - e^2 \sqrt{g}\, \chi^2 \bigr) - 2e^2 \int_{\partial \mathcal{M}} du \, \chi_b^2 (1+e^2 \kappa \epsilon \, \chi_b^2) \sqrt{h}.
\end{equation}
By Gaussian integration on $\chi$, we obtain back precisely the action for 2d Maxwell theory, as expected, coupled to JT gravity in the bulk, namely:
\begin{equation}\label{eq:2dmaxwellbdyterms}
    I \supset \frac{1}{4e^2} \int_{\mathcal{M}} d^2x \, \sqrt{g} \, f_{\mu \nu}f^{\mu \nu} - 2\int_{\partial \mathcal{M}} du \, \chi_b \, a_{u}   - 2e^2 \int_{\partial \mathcal{M}} du \, \chi_b^2 (1+e^2 \kappa \epsilon \, \chi_b^2) \sqrt{h}.
\end{equation}
Using Stokes' theorem, the Maxwell term can be recast as a pure boundary contribution \cite{Blommaert:2018oue}; imposing the Maxwell equations $d \, \hodge f = 0$ then yields
\begin{equation}
    I\Big|_{\text{on-shell}} \supset -\frac{1}{2e^2} \int_{\partial \mathcal{M}} du \, \mathcal{E} \, a_u -2 \int_{\partial \mathcal{M}} du \, \chi_b a_{u}   - 2e^2 \int_{\partial \mathcal{M}} du \, \chi_b^2 (1+e^2 \kappa \epsilon \, \chi_b^2) \sqrt{h}.
\end{equation}
Evaluating \eqref{eq:chiequation} at the boundary, the first two terms can be combined to give:
\begin{equation}
    I\Big|_{\text{on-shell}} \supset  -\int_{\partial \mathcal{M}} du \, \chi_b \, a_{u}   - 2e^2 \int_{\partial \mathcal{M}} du \, \chi_b^2 (1+e^2 \kappa \epsilon \, \chi_b^2) \sqrt{h}.
\end{equation}
Now, we can integrate out fully $\chi_b$ (or equivalently, the electric field $\mathcal{E}$) to get the on-shell boundary action of 2d Maxwell: 
\begin{equation}
    I\Big|_{\text{on-shell}} \supset \frac{1}{2\kappa} \int_{\partial \mathcal{M}} du \, \left(-\frac{a_u}{2e^2\chi_b}-\sqrt{h}\right)^2 = \frac{1}{2\kappa} \int_{\partial \mathcal{M}} du \, \left(\frac{a_u}{\mathcal{E}}-\sqrt{h}\right)^2
\end{equation}
This is precisely of the same form as \eqref{eq:mixedonshellaction}, where we can identify again via field redefinition $A_{u} = -\tfrac{a_u}{2e^2 \chi_b}$, and this gives the free particle on $U(1)$ action upon solving for $\hodge f\big|_{\partial \mathcal{M}} = -2e^2\chi_b=\mathcal{E}$ and Maxwell's equation of motion $d \, \hodge f = 0$, with the solution given by
\begin{equation}
    a_u\big|_{\partial \mathcal{M}} = \frac{\mathcal{E}}{\epsilon} + i \varphi_{M}' \equiv \mathcal{E} \left(\frac{1}{\epsilon} + i \varphi'\right). 
\end{equation}
We can clearly see that the hence obtained Maxwell solution is related to the HT asymptotic solution \eqref{eq:gaugefieldasymptoticsolutionPoincare} via the electric field $\mathcal{E}$, and where $\varphi_M(u) = \mathcal{E} \,  \varphi(u)$ is the Maxwell pure gauge configuration. The corresponding free particle on $U(1)$ action for 2d Maxwell coming from HT$_2$ takes the form 
\begin{equation}
     I\Big|_{\text{on-shell}} \supset - \frac{1}{2\kappa \mathcal{E}^2} \int_{\partial \mathcal{M}} du \, \varphi'^2_M \equiv - \frac{1}{2\kappa} \int_{\partial \mathcal{M}} du \, \varphi'^2.
\end{equation}
Clearly, HT$_2$ gravity and 2d Maxwell theory are tied through the vacuum cosmological constant. Starting from HT$_2$ with mixed boundary conditions and rewriting the theory in Maxwell variables via a field redefinition, we recover the same boundary action, namely the particle on $U(1)$. Moreover, it is not surprising that the factor of $\mathcal{E}^2$ is absent in the HT variables, since this is a characteristic feature of the HT formulation, as discussed in section \ref{sec:2.3}.

Therefore, we have shown that HT$_2$ gravity, equipped with the appropriate boundary conditions, is equivalent to JT gravity coupled to 2d Maxwell theory. From this perspective, it is natural that HT$_2$ gravity with mixed boundary conditions \eqref{eq:mixedbdyconditions} on the fields $(\Lambda_b, A_u)$ yields the effective action of complex SYK at the boundary. Indeed, combining the bulk equivalence with the fact that we can directly reduce 2d Maxwell on a disk to a particle on $U(1)$ living on the boundary of the disk \cite{Blommaert:2018oue} completes the argument: both theories lead to the same boundary dynamics, and hence provide equivalent descriptions in the bulk and on the boundary.

\paragraph{Gauge transformations for field redefinitions.}

Having established the equivalence between 2d Maxwell theory and HT$_2$ gravity explicitly, we now ask whether, after the field redefinition, the new field $A_\mu$ still transforms as a genuine gauge field. To address this, we examine the gauge symmetry of the action under the transformation of $A_\mu$ induced by the original gauge transformation of $a_\mu$.

The 2d Maxwell action given in \eqref{eq:JT2dMaxwell} is gauge invariant under the gauge transformation of $a_\mu$, namely
\begin{equation}
    a_\mu \longmapsto a_\mu + i \partial_\mu \vartheta_1,
\end{equation}
where we take $\vartheta_1$ to be a single-valued function of spacetime, that is, we do not consider large gauge transformations for the time being. The equivalent $\chi^2$-action \eqref{eq:quadraticaction} is also gauge invariant under the same gauge transformation of $a_\mu$. The linear $\Lambda$-action \eqref{eq:lambdaaction} is clearly invariant under the following transformation of $A_\mu$:
\begin{equation}
    A_\mu \longmapsto A_\mu + i \partial_\mu \vartheta_2,
\end{equation}
where we have taken $\vartheta_2$ to be a single-valued function of spacetime. Note that it is gauge invariant modulo boundary transformation. Accordingly, it is the invariance of the respective actions under these transformations that identifies $a_\mu$ and $A_\mu$ as \textit{bona fide} gauge fields.

Now, suppose we induce the transformation of $A_\mu$ from that of the gauge transformation of $a_\mu$, that is
\begin{equation}\label{eq:inducedtrfs}
    a_\mu \longmapsto a_\mu + i \partial_\mu \vartheta_1 \ \ \ \ \ \ \Longrightarrow \ \ \ \ \ \ A_\mu \longmapsto A_\mu - \frac{i}{2e^2 \chi} \partial_\mu \vartheta_1.
\end{equation}
We then check whether the action \eqref{eq:lambdaaction} remains invariant under this transformation induced from $a_\mu$. After transforming the $\Lambda$-action, we find:
\begin{equation}\label{eq:HT2trfs}
    I_{\text{HT$_2$}} \longmapsto I_{\text{HT$_2$}} - 2i \int_{\partial \mathcal{M}} du \, \chi_b \, \vartheta'_1(u).
\end{equation} 
The action is therefore gauge invariant up to a boundary term. We can then consider 2d Maxwell theory, supplemented by boundary terms \eqref{eq:2dmaxwellbdyterms} chosen to be compatible with the HT$_2$ action, and perform a gauge transformation generated by $a_u$. The resulting variation of the action coincides precisely with the HT$_2$ transformation laws \eqref{eq:HT2trfs}. This suggests that the two theories have the same charge at the boundary.

\subsection{HT\texorpdfstring{$_2$}{HT2} gravity as a BF gauge theory}

In this section, we analyse the extended AdS algebra $\reallywidehat{\mathfrak{sl}}(2,\mathbb{R})$ and develop a manifestly decoupled BF formulation of HT$_2$ gravity. Working in the first-order formalism makes both the algebraic extension and the decoupling structure transparent. We begin by setting notation with a brief review of the first-order formulation of JT gravity, then construct the BF reformulation of HT$_2$.

\paragraph{AdS algebra and JT gravity.}

The AdS Lie algebra $\mathfrak{sl}(2,\mathbb{R})$ is described by the generators $P_a$ (translations in $(1+1)$) and $\widetilde{J} = \tfrac{1}{2}\epsilon^{ab} J_{ab}$ (rotation and boost in $(1+1)$). They satisfy the commutation relations
\begin{align}
    \bigl[ P_a,P_b\bigr] = -\lambda \epsilon_{ab} \widetilde{J}, \ \ \ \ \ \bigl[P_a,\widetilde{J}\bigr] = -\epsilon_a^{\ b} P_b,  \ \ \ \ \ \bigl[\widetilde{J},\widetilde{J}\bigr] = 0.
\end{align}
One can consider then an $\mathfrak{sl}(2,\mathbb{R})$-valued connection $\mathbf{A}$ and curvature $\mathbf{F}$, respectively:
\begin{equation}
    \mathbf{A} = e^a P_a + \widetilde{\omega} \widetilde{J},
\end{equation}
\begin{equation}
    \mathbf{F} = d \mathbf{A} + \mathbf{A} \wedge \mathbf{A} = T^a P_a + \bigl( \widetilde{R} - \lambda \widetilde{\Sigma}\bigr) \widetilde{J}.
\end{equation}
Here, $e^a$ is the tetrad and $\widetilde{\omega}=\tfrac{1}{2}\epsilon^{ab}\omega_{ab}$ is the (dual) spin connection. We define the area form as $\widetilde{\Sigma} = \tfrac{1}{2}\,\epsilon_{ab} e^a \wedge e^b$. The curvature of the tetrad is the torsion $T^a = De^a$ and the curvature of the spin connection is denoted by $\widetilde{R} = d\widetilde{\omega}$. Then, we can write JT gravity action as a BF-like theory:
\begin{equation}
    S_{\text{JT}} = \int B_a T^a + \phi \bigl(\widetilde{R} - \lambda \widetilde{\Sigma}\bigr),
\end{equation}
where the Lagrange multiplier fields $B_a$ and $\phi$ ensure that on-shell we have
\begin{equation}
    \mathbf{F} = 0 \ \ \ \Longleftrightarrow \ \ \ T^a = 0, \ \ \ \ \ \ \widetilde{R} = \lambda \widetilde{\Sigma}.
\end{equation}
These are equivalent in the second-order formalism to torsion-free connection and JT gravity equation $R+2 = 0$. 

\paragraph{Extended de Sitter algebra and HT$_2$ gravity.}

We now consider a central extension of the AdS algebra by the abelian $U(1)$ Lie group. The resulting ``extended AdS'' algebra is the direct sum $\mathfrak{sl}(2,\mathbb{R}) \oplus \mathfrak{u}(1)$ with generators $P_a$, $\widetilde{J}$, and $I$ (central element). The non-vanishing commutation relations are
\begin{align}
    \bigl[ P_a,P_b\bigr] &= -\lambda \epsilon_{ab} \widetilde{J} + \epsilon_{ab} I, \qquad \bigl[P_a,\widetilde{J}\bigr] = -\epsilon_a^{\ b} P_b.
\end{align}
We can then consider a connection and curvature valued in the extended algebra: the connection becomes
\begin{equation}
    \mathbf{A} = e^a P_a + \widetilde{\omega} \widetilde{J} + A I,
\end{equation}
and the curvature is
\begin{equation}
    \mathbf{F} = T^a P_a + \bigl(\widetilde{R} - \lambda \widetilde{\Sigma}\bigr) \widetilde{J} + \bigl(F+\widetilde{\Sigma}\bigr) I.
\end{equation}
Here, we used the abelian curvature  $F=dA$. We can now consider an action akin to JT gravity that is BF-like:
\begin{equation}
    S_1 = \int B_a T^a + \phi \bigl(\widetilde{R} - \lambda \widetilde{\Sigma}\bigr) - \Lambda \bigl(F+\widetilde{\Sigma}\bigr),
\end{equation}
where the Lagrange multiplier fields $B_a$, $\phi$, and $\Lambda$ ensure that on-shell we have
\begin{equation}
    \mathbf{F} = 0 \ \ \ \Longleftrightarrow \ \ \ T^a = 0, \ \ \ \ \ \ \widetilde{R} = \lambda \widetilde{\Sigma},  \ \ \ \ \ \ F = -\widetilde{\Sigma}.
\end{equation}
Because we recover the same classical dynamics as JT gravity supplemented by the unimodular HT condition $F = -\widetilde{\Sigma}$, the action $S_1$ is described by HT$_2$ gravity.

There exists an ambiguity in two dimensions for the generator $\widetilde{J}$ since the group of rotations on the plane is $\mathfrak{so}(1,1)$ which is isomorphic to $\mathfrak{u}(1)$ (algebraic equivalence, not group level isomorphism). Therefore, we can  combine the generators $\widetilde{J}$ and $I$ into a new abelian generator $\widetilde{L}$ by writing
\begin{equation}
    \widetilde{L} := \widetilde{J} - \frac{I}{\lambda}.
\end{equation}
We can then re-write the non-vanishing commutation relations for the extended AdS algebra in a way that $\mathfrak{sl}(2,\mathbb{R})$ and $\mathfrak{u}(1)$ are ``decoupled''\footnote{Consider replacing the central element $I \rightarrow \mathcal{B}I$ (see \cite{Gonzalez:2018enk}. The translation generators commutation relation changes to:
\begin{align}
    \bigl[ P_a,P_b\bigr] &= -\lambda \epsilon_{ab} \widetilde{J} + \mathcal{B}\, \epsilon_{ab} I.
\end{align}
This can also be obtained by performing a shift of the generator $\widetilde{L} \rightarrow \widetilde{L} - \tfrac{\mathcal{B} I}{\lambda}$ in the decoupled algebra. The addition of such a term enforces that one can take the limit $\mathcal{B} \rightarrow 0$.}:
\begin{align}
    \bigl[ P_a,P_b\bigr] &= -\lambda \epsilon_{ab} \widetilde{L}, \qquad \bigl[P_a,\widetilde{L}\bigr] = -\epsilon_a^{\ b} P_b.
\end{align}
The connection takes the form
\begin{equation}
    \mathbf{A} = e^a P_a + \widetilde{\omega} \widetilde{L} + \mybar{0.9}{2pt}{A}I,
\end{equation}
where $\mybar{0.9}{2pt}{A} = A + \tfrac{\widetilde{\omega}}{\lambda}$. The associated curvature two-form is
\begin{equation}
    \mathbf{F} = T^a P_a + \bigl(\widetilde{R} - \lambda \widetilde{\Sigma}\bigr) \widetilde{L} + \mybar{0.9}{2pt}{F}I,
\end{equation}
with the abelian curvature $\mybar{0.9}{2pt}{F} = F + \tfrac{\widetilde{R}}{\lambda}$. We can now consider once more a BF-like action
\begin{equation}
    S_2= \int B_a T^a +\mybar{0.8}{1pt}{\phi} \bigl(\widetilde{R} - \lambda \widetilde{\Sigma}\bigr) - \mybar{0.7}{1pt}{\Lambda} \mybar{0.9}{2pt}{F},
\end{equation}
where the Lagrange multiplier fields $B_a$, $\mybar{0.8}{1pt}{\phi}$, and $\mybar{0.7}{1pt}{\Lambda}$ ensures that on-shell we have
\begin{equation}
    \mathbf{F} = 0 \ \ \ \Longleftrightarrow \ \ \ T^a = 0, \ \ \ \ \ \ \widetilde{R} = \lambda \widetilde{\Sigma},  \ \ \ \ \ \ \mybar{0.9}{2pt}{F} = 0.
\end{equation}
It is rather interesting that this action describes an abelian BF gauge theory in a completely decoupled way from JT gravity.

The question now is whether we can relate both actions $S_1$ and $S_2$. For that, consider expressing the decoupled variables $\mybar{0.8}{1pt}{\phi}$ and $\mybar{0.7}{1pt}{\Lambda}$ as linear combinations of $\phi$ and $\Lambda$:
\begin{equation}
    \mybar{0.8}{1pt}{\phi} (\phi, \Lambda) = \alpha_1 \phi + \alpha_2 \Lambda, \ \ \ \ \ \ \mybar{0.7}{1pt}{\Lambda}(\phi, \Lambda) = \beta_1 \phi + \beta_2 \Lambda.
\end{equation}
We can then expand $S_2$ and find
\begin{align}%\hspace{-1 cm}
    S_2 &= \int B_a T^a + \bigl(\alpha_1 \phi + \alpha_2 \Lambda - \lambda^{-1} \beta_1 \phi - \lambda^{-1} \beta_2 \Lambda\bigr) \widetilde{R} - \lambda \alpha_1 \phi \widetilde{\Sigma} - \lambda \alpha_2 \Lambda \widetilde{\Sigma} - \beta_1 \phi F - \beta_2 \Lambda F \nonumber\\
    &= \int B_a T^a + \phi\bigl[(\alpha_1 - \lambda^{-1} \beta_1) \widetilde{R} - \alpha_1 \lambda \widetilde{\Sigma} - \beta_1 F\bigr] -\Lambda \bigl[(-\alpha_2 + \lambda^{-1} \beta_2) \widetilde{R} + \alpha_2 \lambda \widetilde{\Sigma} + \beta_2 F\bigr].
\end{align}
Hence, we can see that to recover $S_1$ from $S_2$, we need to impose the following conditions on the coefficients:
\begin{equation}
    \alpha_1 = 1, \ \ \ \beta_2 = 1, \ \ \ \text{and} \ \ \ \alpha_2 = \frac{\beta_2}{\lambda}, \ \ \ \beta_1 = 0.
\end{equation}
In other words, from the relations
\begin{equation}
    \mybar{0.8}{1pt}{\phi} (\phi, \Lambda) = \phi + \frac{\Lambda}{\lambda}, \ \ \ \ \ \ \ \ \mybar{0.7}{1pt}{\Lambda}(\phi, \Lambda) = \Lambda, \ \ \ \ \ \ \ \ \ \mybar{0.9}{2pt}{A} = A + \frac{\widetilde{\omega}}{\lambda},
\end{equation}
HT$_2$ gravity is equivalently described by JT gravity coupled to $\mathfrak{u}(1)$ BF theory, i.e.:
\begin{align}
    S_{\text{HT}_2} = \int B_a T^a + \phi \bigl(\widetilde{R} - \lambda \widetilde{\Sigma}\bigr) - \Lambda \bigl(F+\widetilde{\Sigma}\bigr) = \int B_a T^a +\mybar{0.8}{1pt}{\phi} \bigl(\widetilde{R} - \lambda \widetilde{\Sigma}\bigr) - \mybar{0.7}{1pt}{\Lambda} \mybar{0.9}{2pt}{F}.
\end{align}
Therefore, a deep interplay between the extended AdS algebra $\reallywidehat{\mathfrak{sl}(2,\mathbb{R})}$ and the realization of HT$_2$ gravity as JT $+$ abelian BF theory exist.

Note that the decoupled theory features a shifted dilaton $\mybar{0.8}{1pt}{\phi}$. In contrast to the shifted JT model, where a constant shift $\phi \to \phi + \tfrac{\Lambda}{\lambda}$ (with $\Lambda$ constant) yields pure JT gravity, here one must also shift the abelian gauge field, $A \to A + \tfrac{\widetilde{\omega}}{\lambda}$. With this simultaneous redefinition, the theory factorises into JT gravity plus a $\mathfrak{u}(1)$ BF sector, with no residual bulk coupling between them. 

It is worth emphasising that, although the extended AdS algebra admits a decoupled action yielding JT gravity plus a $U(1)$ BF sector, the situation differs for the $\reallywidehat{\text{CGHS}}$ model: while it has a BF formulation based on the Maxwell algebra, it does not factorise into a flat JT sector and a $U(1)$ BF theory.

\subsection{Towards embedding HT\texorpdfstring{$_2$}{HT2} in higher-dimensions}

Building on the machinery developed in this section, we can attempt to embed HT$_2$ gravity into a four-dimensional black-hole setting. Concretely, an emergent $U(1)$ sector can arise either from a dimensional reduction of the near-horizon geometry of charged black holes, such as the Reissner-Nordstr\"{o}m solution in various backgrounds, or, more generally, from reducing on an internal space, in which case the gauge group is inherited from the isometries of the compact manifold \cite{Iliesiu:2019lfc}.

A convenient starting point is the general Euclidean action describing JT gravity coupled to two-dimensional Yang-Mills. For an arbitrary gauge group $G$ and dilaton-dependent coupling $e(\phi)$, the action is given by \cite{Iliesiu:2019lfc}:
\begin{align}
    I = &-\frac{\phi_0}{2} \int_{\mathcal{M}} d^2x \, \sqrt{g} \, R - \frac{1}{2} \int_{\mathcal{M}} d^2x \, \sqrt{g} \, \phi(R+2) + \int_{\mathcal{M}} d^2x  \, \frac{ \sqrt{g}}{4e^2(\phi)} \,\mathrm{Tr} \, (f_{\mu \nu} f^{\mu \nu}) + I_{\text{bdy}}.
\end{align}
Let us define the following change of variables for the coupling constants to get an effective dilaton-dependent coupling
\begin{equation}
    \frac{1}{e(\phi)} = \frac{1}{e^2} + \frac{\phi_0 + \phi}{e_{\phi}^2 \phi_0} = \frac{e^2\left(1+\frac{\phi}{\phi_0}\right) + e_\phi^2}{e^2 e_\phi^2}.
\end{equation}
One can then introduce a scalar field $\chi$, viewed as a $G$-adjoint-valued $0$-form, and rewrite the theory in the quadratic form used in the previous subsections, now with a dilaton-dependent effective coupling:
\begin{align}
    I \supset - \int_{\mathcal{M}} d^2x \, \mathrm{Tr} \, (\chi \, \varepsilon^{\mu \nu} f_{\mu \nu}) - 2 \int_{\mathcal{M}} d^2x \, \sqrt{g} \, e(\phi) \mathrm{Tr}\,(\chi^2) + I_{\text{bdy}}.
\end{align}

In the context of higher-dimensional black holes, we are essentially interested in describing the near-horizon region close to extremality; thus, we take the weak dilaton limit $\phi \ll \phi_0$. It is therefore convenient to parametrise the effective couplings as follows:
\begin{equation}
    e(\phi) \equiv \widetilde{e} - \widetilde{e}_{\phi} \phi, \qquad \widetilde{e} := \frac{e^2 e_\phi^2}{e^2 + e_\phi^2}, \qquad \widetilde{e}_{\phi} := \frac{e^4 e_\phi^2}{(e^2 + e_\phi^2)^2 \phi_0},
\end{equation}
where we have defined the near-extremal couplings $\widetilde{e}$ and $\widetilde{e}_{\phi}$.
We can then rewrite the quadratic action as
\begin{equation}
    I \supset - \int_{\mathcal{M}} d^2x \, \mathrm{Tr} \, (\chi \, \varepsilon^{\mu \nu} f_{\mu \nu}) - 2 \int_{\mathcal{M}} d^2x \, \sqrt{g} \, (\widetilde{e} - \widetilde{e}_{\phi} \phi) \mathrm{Tr} \, (\chi^2) + I_{\text{bdy}}.
\end{equation}
It is of interest to check whether this action with an arbitrary dilaton-dependent coupling, and for the choice of the gauge group $G$ to be $U(1)$, differs in any way from HT$_2$ gravity after field redefinition and through the central extension mechanism explored in previous subsections. We also note that the electric field is dilaton-dependent in this context $\mathcal{E}\sim e(\phi)\chi$.

\paragraph{A minimally extended HT$_2$ gravity.}

We shall first comment on the weak coupling limit, where $\widetilde{e}, \widetilde{e}_{\phi} \rightarrow 0$. This is particularly relevant as this yields a pure BF action \cite{Witten:1991we,Iliesiu:2019lfc}:
\begin{equation}
    I \supset - \int_{\mathcal{M}} d^2x \, \chi \, \varepsilon^{\mu \nu} f_{\mu \nu} + I_{\text{bdy}}.
\end{equation}
Combined with the JT gravity sector, this yields the extended AdS algebra $\mathfrak{sl}(2,\mathbb{R}) \oplus \mathfrak{u}(1)$, which is a trivial central extension of the anti-de Sitter algebra. One can hence rewrite the action as HT$_2$ gravity by performing the simultaneous shifts of the dilaton and gauge field, as discussed in the previous subsection. The boundary theory, as expected, is related to the particle on the $SL(2,\mathbb{R}) \times U(1)$ manifold and therefore related to the low-energy effective action of the complex SYK model, as argued in this paper.

Let us now consider the non-weak coupling regime and perform the field redefinitions on the quadratic action 
\begin{equation}
    \Lambda = -2 \chi^2, \, \, \, \, \, \, \, \, \, \, \, \, \, \,  A_{\mu} = - \frac{a_{\mu}}{2 \chi}.
\end{equation}
It is straightforward to then show that this yields the action in the $(\Lambda,A_{\mu})$ variables given by:
\begin{equation}
    I = -\frac{1}{2} \int_{\mathcal{M}} d^2x \, \sqrt{g} \, \phi(R+2) + \int_{\mathcal{M}} d^2x \, \Lambda \bigl(e(\phi) \sqrt{g} - \varepsilon^{\mu \nu} \partial_{\mu} A_{\nu} \bigr).
\end{equation}
This is clearly a minimal extension of HT$_2$ gravity action, now with the square-root of the determinant of the metric tensor multiplied by the dilaton-dependent coupling constant. This minimal modification has, however, drastic consequences for both the JT gravity and HT sectors of the theory. To see this, consider the metric and dilaton equations of motion:
\begin{equation}
    R+2 -2 \Lambda e'(\phi) = 0, \, \, \, \, \, \, \, \,\ \,\,\,\,\,\,\, \bigl(\nabla_{\mu} \partial_{\nu} - g_{\mu \nu} \, \Box\bigr) \phi = \bigl(\phi - \Lambda e(\phi)\bigr)g_{\mu \nu}.
\end{equation}
Clearly, the JT gravity equations are modified, as the dilaton-dependent coupling is changing the background geometry: Writing $e(\phi) = \widetilde{e} - \widetilde{e}_{\phi}\phi$, we get:
\begin{equation}
    R+2(1 + \widetilde{e}_{\phi}\Lambda) = 0, \,\,\,\,\,\,\,\,\,\,\,\,\,\,\,\, \bigl(\nabla_{\mu} \partial_{\nu} - g_{\mu \nu} \, \Box\bigr) \phi = \bigl[(1+\widetilde{e}_{\phi}\Lambda)\phi - \widetilde{e}\Lambda \bigr]g_{\mu \nu}.
\end{equation}
We find that the vacuum cosmological constant \(\Lambda\) induces an effective cosmological constant
\begin{equation}
    \lambda_{\text{eff}}(\Lambda) = -1 - \widetilde e_{\phi}\,\Lambda,
\end{equation}
so that the curvature constraint becomes $R-2\lambda_{\text{eff}}(\Lambda)=0$. In the perturbative regime $|\widetilde e_{\phi}\Lambda|\ll 1$, both the dilaton and metric equations reproduce the usual shifted-JT on-shell relations. (In particular, for $\Lambda>0$ one obtains an $AdS_2$ background.) This makes clear that $\Lambda$ enters the theory in two distinct ways: through $\widetilde e$ it controls the vacuum-energy sector, while through $\widetilde e_{\phi}$ it backreacts on the background geometry. In other words, the dependence of $\lambda_{\text{eff}}$ on $\Lambda$ arises explicitly from the dilaton coupling $\widetilde e_{\phi}$.

On the other hand, variation with respect to $\Lambda$ and $A_{\mu}$ results in the on-shell expressions:
\begin{align}
    e(\phi)\sqrt{g} &= \varepsilon^{\mu \nu} \partial_\mu A_\nu, \qquad \partial_{\mu} \Lambda = 0. 
\end{align}
Hence, the unimodular HT constraint is modified accordingly. This admits a natural interpretation once we recall features of two-dimensional Yang--Mills theory. Consider fixing the dilaton and metric fields as backgrounds, with near-extremal coupling $e(\phi) = \widetilde{e} - \widetilde{e}_\phi \phi$  as an arbitrary source for $\chi^2$. Then, it is well known that the source can be absorbed into the two-dimensional area integral: the fact that the theory is invariant under local area-preserving diffeomorphisms defines the dimensionless quantity $\mathcal{A} = \int_{\mathcal{M}} d^2x \, \sqrt{g} \, e(\phi)$\footnote{That being said, we can rewrite the HT constraint equation via a Weyl rescaled metric
\begin{equation}
    \widehat{g}_{\mu \nu} = \bigl(\widetilde{e} - \widetilde{e}_\phi \phi\bigr)g_{\mu \nu} \ \ \ \ \ \ \ \ \ \Longrightarrow \ \ \ \ \ \ \ \ \ \sqrt{\widehat{g}} = \bigl(\widetilde{e} - \widetilde{e}_\phi \phi\bigr)\sqrt{g}.
\end{equation}
Therefore, one can view the HT constraint equation as relating the determinant of the Weyl rescaled metric to the field strength, that is
\begin{equation}
    \sqrt{\widehat{g}} = \varepsilon^{\mu \nu} \partial_{\mu} A_\nu.
\end{equation}
This does not change the notion of spacetime volume being a difference of unimodular time flux; it just shifts the perspective to a dimensionless area density.}.

In line with the results of the previous section, one can anticipate an effective boundary description for this theory. Indeed, we expect a closely analogous structure---already explored in earlier work by \cite{Iliesiu:2019lfc}---in which the boundary dynamics is again that of a particle on $U(1)$, but with a modified defect potential.

\paragraph{A new algebra?}

It is tempting to write the modified HT$_2$ action in first-order formalism as
\begin{equation}
    S = \int B_a T^a + \phi \bigl(\widetilde{R} - \lambda \widetilde{\Sigma}\bigr) - \Lambda \bigl(F + e(\phi) \widetilde{\Sigma}\bigr).
\end{equation}
Using the effective cosmological constant $\lambda_{\text{eff}}(\Lambda) = \lambda - \widetilde{e}_\phi \Lambda$ as a variable, we can rewrite the action above as follows:
\begin{equation}
    S = \int B_a T^a + \phi \bigl(\widetilde{R} - \lambda_{\text{eff}}(\Lambda) \widetilde{\Sigma}\bigr) - \Lambda \bigl(F + \widetilde{e} \, \widetilde{\Sigma}\bigr).
\end{equation}
Notice that this action is not a BF gauge theory due to the presence of the $\Lambda$-dependent effective cosmological constant, which spoils the Lagrange multiplier aspect of the vacuum cosmological constant in the standard formulation of HT$_2$ gravity. Regardless, it is plausible to look for an algebraic construction of such an action. For this, consider the following algebra of commutation relations:
\begin{align}
    \bigl[ P_a,P_b\bigr] &= -\lambda_{\text{eff}}(\Lambda) \, \epsilon_{ab} \widetilde{J} + \widetilde{e} \, \epsilon_{ab} I, \quad \bigl[P_a,\widetilde{J}\bigr] = -\epsilon_a^{\ b} P_b,  \quad \bigl[\widetilde{J},\widetilde{J}\bigr] = 0.
\end{align}
The connection valued in the algebra takes the form
\begin{equation}
    \mathbf{A} = e^a P_a + \widetilde{\omega} \widetilde{J} + A I,
\end{equation}
and the associated curvature is therefore
\begin{equation}
    \mathbf{F} = T^a P_a + \bigl(\widetilde{R} - \lambda_{\text{eff}} \, \widetilde{\Sigma}\bigr) \widetilde{J} + \bigl(F + \widetilde{e} \, \widetilde{\Sigma}\bigr) I.
\end{equation}
Introducing the 0-form fields $\mathbf{B} = (B_a,-\lambda_{\text{eff}} \, \phi - \Lambda,\phi + \Lambda)$, we can then consider a BF-like action that reproduces the correct first-order action introduced above, namely\footnote{We have used the following inner products:
\begin{align}
    \langle P_a,P_b\rangle = \eta_{ab}, \ \ \ \ \ \ \langle\widetilde{J},\widetilde{J}\rangle = \langle\widetilde{J},I\rangle= \frac{1}{1-\lambda_{\text{eff}}}, \ \ \ \ \ \ \langle I,I \rangle = \frac{\lambda_{\text{eff}}}{1-\lambda_{\text{eff}}}.
\end{align}}:
\begin{align}
    S = \int \langle \mathbf{B} , \mathbf{F} \rangle = \int B_a T^a + \phi\bigl(\widetilde{R} - \lambda_{\text{eff}} \, \widetilde{\Sigma}\bigr) - \Lambda \bigl(F + \widetilde{e} \,  \widetilde{\Sigma}\bigr).
\end{align}
Again, we stress that this is not a BF due to the fact that the effective cosmological constant is $\Lambda$-dependent. What is meant by ``BF-like'' is the way the $\mathbf{B}$ field enters in the inner product between $\mathbf{F}$ at the level of the action.

\section{Unimodular time on the Euclidean disk}\label{sec:4}

Unimodular time plays a central role in the canonical quantisation of gravity: it is canonically conjugate to the vacuum cosmological constant, with Poisson bracket $\{T_{\Lambda},\Lambda\}_{\text{P.B.}}=1$, so the Wheeler-DeWitt constraint becomes a Schr\"odinger-type evolution equation in $T_{\Lambda}$. In what follows, we do not revisit the explicit quantisation procedure (see \cite{Alexandre:2025rgx} for a discussion of the quantisation of de Sitter HT$_2$ gravity); instead, we examine the structure and boundary implications of unimodular time. Building on the previous two sections, where the low-energy complex SYK model reproduces the correct boundary dynamics of Euclidean HT$_2$ gravity, as expected from its equivalence to JT gravity coupled to 2d Maxwell and BF gauge theories, we now focus on the holographic role of unimodular time at the boundary.

In the HT$_2$ gravity framework, assuming an appropriate slicing of the manifold as $\mathcal{M} \cong \Sigma \times \mathbb{R}$, and using $\mathcal{T}^\mu = \varepsilon^{\mu \nu} A_\nu$ as the densitised dual of $A_\mu$, unimodular time $T_{\Lambda}$ on the constant slice $\Sigma$ is defined as the flux integral of the gauge field $A_\mu$ through the surface, namely:
\begin{equation}
    T_{\Lambda}(\Sigma) := \int_{\Sigma} du \, n_{\mu} \mathcal{T}^{\mu} = \int_\Sigma dx^\mu\,A_{\mu} \equiv \mathrm{Flux}(A|_{\Sigma}),
\end{equation}
where we parametrised the surface by coordinates $x^{\mu}(u)$. 

The notion of unimodular time is inherently foliation-dependent: it is defined only after specifying a hypersurface $\Sigma$ and a slicing $\mathcal{M}\cong \Sigma\times \mathbb{R}$. In Euclidean signature, there is no causal structure selecting a preferred notion of time, so one may choose the foliation that is best adapted to the physical interpretation. In the present context, we will adopt a radial $\rho$ slicing, with metric
\begin{equation}
    ds^2 \;=\; r_h^2\sinh^2(\rho)\,d\tau^2 + d\rho^2.
\end{equation}
There are several motivations for this choice. First, it aligns naturally with holographic intuition: radial evolution is the Euclidean analogue of ``bulk time'' in radial quantisation, where one regards the bulk Hilbert space as generated by evolution away from the asymptotic boundary. Second, this is precisely the kind of slicing one wants for unimodular time: the decomposition into constant-$\rho$ hypersurfaces defines a minisuperspace metric.\\
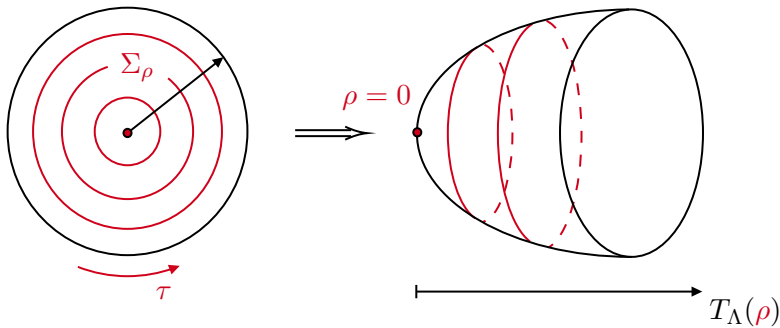
\begin{figure}[!ht]
    \centering

\tikzset{every picture/.style={line width=0.75pt}} %set default line width to 0.75pt        

\begin{tikzpicture}[x=0.75pt,y=0.75pt,yscale=-1,xscale=1]
%uncomment if require: \path (0,202); %set diagram left start at 0, and has height of 202

%Shape: Arc [id:dp2521475994736857] 
\draw  [draw opacity=0][dash pattern={on 4.5pt off 4.5pt}] (439.99,16.8) .. controls (449.9,21.33) and (457.46,45.25) .. (457.46,74.08) .. controls (457.46,102.05) and (450.35,125.4) .. (440.87,130.9) -- (436.58,74.08) -- cycle ; \draw  [color={rgb, 255:red, 208; green, 2; blue, 27 }  ,draw opacity=1 ][dash pattern={on 4.5pt off 4.5pt}] (439.99,16.8) .. controls (449.9,21.33) and (457.46,45.25) .. (457.46,74.08) .. controls (457.46,102.05) and (450.35,125.4) .. (440.87,130.9) ;  
%Shape: Arc [id:dp3899600273022359] 
\draw  [draw opacity=0][dash pattern={on 4.5pt off 4.5pt}] (409.01,28.88) .. controls (416.89,31.82) and (422.97,50.57) .. (422.97,73.25) .. controls (422.97,96.68) and (416.49,115.91) .. (408.23,117.86) -- (406.84,73.25) -- cycle ; \draw  [color={rgb, 255:red, 208; green, 2; blue, 27 }  ,draw opacity=1 ][dash pattern={on 4.5pt off 4.5pt}] (409.01,28.88) .. controls (416.89,31.82) and (422.97,50.57) .. (422.97,73.25) .. controls (422.97,96.68) and (416.49,115.91) .. (408.23,117.86) ;  
%Shape: Arc [id:dp568259821542489] 
\draw  [draw opacity=0] (431.69,127.76) .. controls (422.43,120.66) and (415.7,99.05) .. (415.7,73.49) .. controls (415.7,46.81) and (423.03,24.43) .. (432.92,18.37) -- (438.3,73.49) -- cycle ; \draw  [color={rgb, 255:red, 208; green, 2; blue, 27 }  ,draw opacity=1 ] (431.69,127.76) .. controls (422.43,120.66) and (415.7,99.05) .. (415.7,73.49) .. controls (415.7,46.81) and (423.03,24.43) .. (432.92,18.37) ;  
%Shape: Arc [id:dp0634591216592858] 
\draw  [draw opacity=0] (404.53,116.18) .. controls (396.56,110.87) and (390.72,93.71) .. (390.72,73.36) .. controls (390.72,52.19) and (397.04,34.47) .. (405.51,29.96) -- (409.75,73.36) -- cycle ; \draw  [color={rgb, 255:red, 208; green, 2; blue, 27 }  ,draw opacity=1 ] (404.53,116.18) .. controls (396.56,110.87) and (390.72,93.71) .. (390.72,73.36) .. controls (390.72,52.19) and (397.04,34.47) .. (405.51,29.96) ;  
%Shape: Arc [id:dp7418778998079484] 
\draw  [draw opacity=0] (482.41,135.53) .. controls (482.39,135.53) and (482.37,135.53) .. (482.34,135.53) .. controls (423.22,135.53) and (375.29,107.66) .. (375.29,73.29) .. controls (375.29,39.14) and (422.6,11.41) .. (481.19,11.06) -- (482.34,73.29) -- cycle ; \draw   (482.41,135.53) .. controls (482.39,135.53) and (482.37,135.53) .. (482.34,135.53) .. controls (423.22,135.53) and (375.29,107.66) .. (375.29,73.29) .. controls (375.29,39.14) and (422.6,11.41) .. (481.19,11.06) ;  
%Shape: Ellipse [id:dp12404373601738061] 
\draw   (482.41,11.05) .. controls (502.12,11.05) and (518.09,38.92) .. (518.09,73.29) .. controls (518.09,107.66) and (502.12,135.53) .. (482.41,135.53) .. controls (462.7,135.53) and (446.72,107.66) .. (446.72,73.29) .. controls (446.72,38.92) and (462.7,11.05) .. (482.41,11.05) -- cycle ;
%Straight Lines [id:da013388507883112633] 
\draw    (374.99,152.98) -- (514.74,152.98) ;
\draw [shift={(517.74,152.98)}, rotate = 180] [fill={rgb, 255:red, 0; green, 0; blue, 0 }  ][line width=0.08]  [draw opacity=0] (5.36,-2.57) -- (0,0) -- (5.36,2.57) -- cycle    ;
\draw [shift={(374.99,152.98)}, rotate = 180] [color={rgb, 255:red, 0; green, 0; blue, 0 }  ][line width=0.75]    (0,3.35) -- (0,-3.35)   ;
%Shape: Ellipse [id:dp36023030345007134] 
\draw   (170.98,72.42) .. controls (170.98,38.05) and (197.77,10.18) .. (230.83,10.18) .. controls (263.89,10.18) and (290.69,38.05) .. (290.69,72.42) .. controls (290.69,106.79) and (263.89,134.66) .. (230.83,134.66) .. controls (197.77,134.66) and (170.98,106.79) .. (170.98,72.42) -- cycle ;
%Shape: Ellipse [id:dp0845284811055601] 
\draw  [color={rgb, 255:red, 208; green, 2; blue, 27 }  ,draw opacity=1 ] (183.69,72.42) .. controls (183.69,45.35) and (204.8,23.41) .. (230.83,23.41) .. controls (256.86,23.41) and (277.97,45.35) .. (277.97,72.42) .. controls (277.97,99.49) and (256.86,121.43) .. (230.83,121.43) .. controls (204.8,121.43) and (183.69,99.49) .. (183.69,72.42) -- cycle ;
%Shape: Arc [id:dp938178268130192] 
\draw  [draw opacity=0] (251.09,45.68) .. controls (258.69,51.91) and (263.57,61.57) .. (263.57,72.42) .. controls (263.57,91.22) and (248.91,106.46) .. (230.83,106.46) .. controls (212.75,106.46) and (198.1,91.22) .. (198.1,72.42) .. controls (198.1,57.2) and (207.71,44.31) .. (220.97,39.95) -- (230.83,72.42) -- cycle ; \draw  [color={rgb, 255:red, 208; green, 2; blue, 27 }  ,draw opacity=1 ] (251.09,45.68) .. controls (258.69,51.91) and (263.57,61.57) .. (263.57,72.42) .. controls (263.57,91.22) and (248.91,106.46) .. (230.83,106.46) .. controls (212.75,106.46) and (198.1,91.22) .. (198.1,72.42) .. controls (198.1,57.2) and (207.71,44.31) .. (220.97,39.95) ;  
%Shape: Ellipse [id:dp020869598185732197] 
\draw  [color={rgb, 255:red, 208; green, 2; blue, 27 }  ,draw opacity=1 ] (214.46,72.42) .. controls (214.46,63.02) and (221.79,55.4) .. (230.83,55.4) .. controls (239.87,55.4) and (247.2,63.02) .. (247.2,72.42) .. controls (247.2,81.82) and (239.87,89.44) .. (230.83,89.44) .. controls (221.79,89.44) and (214.46,81.82) .. (214.46,72.42) -- cycle ;
%Straight Lines [id:da27906025277485336] 
\draw    (230.83,73.2) -- (276.37,36.94) ;
\draw [shift={(278.72,35.08)}, rotate = 141.48] [fill={rgb, 255:red, 0; green, 0; blue, 0 }  ][line width=0.08]  [draw opacity=0] (5.36,-2.57) -- (0,0) -- (5.36,2.57) -- cycle    ;
%Straight Lines [id:da9103547325331859] 
\draw    (314.51,71.79) -- (342.42,71.79)(314.51,74.79) -- (342.42,74.79) ;
\draw [shift={(350.42,73.29)}, rotate = 180] [color={rgb, 255:red, 0; green, 0; blue, 0 }  ][line width=0.75]    (10.93,-3.29) .. controls (6.95,-1.4) and (3.31,-0.3) .. (0,0) .. controls (3.31,0.3) and (6.95,1.4) .. (10.93,3.29)   ;
%Shape: Ellipse [id:dp7732111716395738] 
\draw  [fill={rgb, 255:red, 208; green, 2; blue, 27 }  ,fill opacity=1 ] (373.49,72.88) .. controls (373.49,71.73) and (374.38,70.8) .. (375.48,70.8) .. controls (376.59,70.8) and (377.48,71.73) .. (377.48,72.88) .. controls (377.48,74.02) and (376.59,74.95) .. (375.48,74.95) .. controls (374.38,74.95) and (373.49,74.02) .. (373.49,72.88) -- cycle ;
%Shape: Arc [id:dp2329111208021949] 
\draw  [draw opacity=0] (256.9,140.02) .. controls (248.85,143.39) and (240.05,145.24) .. (230.83,145.24) .. controls (222.11,145.24) and (213.76,143.58) .. (206.06,140.55) -- (230.83,72.42) -- cycle ; \draw [color={rgb, 255:red, 208; green, 2; blue, 27 }  ,draw opacity=1 ]   (254.1,141.12) .. controls (246.82,143.79) and (238.99,145.24) .. (230.83,145.24) .. controls (222.11,145.24) and (213.76,143.58) .. (206.06,140.55) ;  \draw [shift={(256.9,140.02)}, rotate = 159.12] [fill={rgb, 255:red, 208; green, 2; blue, 27 }  ,fill opacity=1 ][line width=0.08]  [draw opacity=0] (5.36,-2.57) -- (0,0) -- (5.36,2.57) -- cycle    ;
%Shape: Ellipse [id:dp3262314381870173] 
\draw  [fill={rgb, 255:red, 208; green, 2; blue, 27 }  ,fill opacity=1 ] (228.84,73.2) .. controls (228.84,72.05) and (229.73,71.12) .. (230.83,71.12) .. controls (231.93,71.12) and (232.83,72.05) .. (232.83,73.2) .. controls (232.83,74.34) and (231.93,75.27) .. (230.83,75.27) .. controls (229.73,75.27) and (228.84,74.34) .. (228.84,73.2) -- cycle ;

% Text Node
\draw (539.4,163.15) node    {$T_{\Lambda }(\textcolor[rgb]{0.82,0.01,0.11}{\rho })$};
% Text Node
\draw (355.17,55.43) node    {$\textcolor[rgb]{0.82,0.01,0.11}{\rho =0}$};
% Text Node
\draw (249.05,153.58) node  [color={rgb, 255:red, 208; green, 2; blue, 27 }  ,opacity=1 ]  {$\textcolor[rgb]{0.82,0.01,0.11}{\tau }$};
% Text Node
\draw (235.59,40.16) node    {$\textcolor[rgb]{0.82,0.01,0.11}{\Sigma _{\rho }}$};

\end{tikzpicture}
    \caption{Left: the disk geometry, with constant-$\rho$ hypersurfaces $\Sigma_\rho$ highlighted in red. Right: the cigar geometry, in which unimodular time induces a preferred foliation, here taken to be the constant-$\rho$ slicing.}
    %\label{fig:placeholder}
\end{figure}\\
Hence, for the Euclidean disk, we take bulk unimodular time to be:
\begin{equation}\label{eq:bulkunimodulartime}
    T_{\Lambda}(\rho) =  \int_{\Sigma_{\rho}} d\tau \, A_\tau.
\end{equation}
The black hole metric \eqref{eq:bhmetric}, written in coordinates $(\tau,r)$, can also be used to define unimodular time, now with constant-$r$ slices. We will use these coordinates below for convenience.

As discussed earlier, unimodular time and the vacuum cosmological constant form a canonical pair in the bulk. It is natural to ask how this relationship imprints itself on the boundary. In Section \ref{sec:2}, we derived the following boundary theory:
\begin{equation}
I^{\text{mixed}}_{\text{bdy}}\Big|_{\text{on-shell}} = - \int_{\partial \mathcal{M}} du \, \sqrt{h} \, \phi_b (K-1) - \int_{\partial \mathcal{M}} du \, \Lambda_b \, i\varphi' - \frac{\kappa \epsilon}{2} \int_{\partial \mathcal{M}} du \, \sqrt{h} \, \Lambda_b^2.
\end{equation}
At this point, following the discussion in Section \ref{sec:3}, we will see that this action should be manifestly equivalent, up to a field redefinition, to the HT gravity action in $(0+1)$ dimensions:
\begin{equation}
I^{\text{mixed}}_{\text{bdy}}\Big|_{\text{on-shell}} = - \int_{\partial \mathcal{M}} du \, \sqrt{h} \, \phi_b (K-1) - \int_{\partial \mathcal{M}} du \, \Theta\,(i \Psi' - \sqrt{h}).
\end{equation}
This provides a natural setting for defining a ``boundary unimodular time'' that is intrinsic to the boundary theory. We shall further see how this relates to the bulk unimodular time evaluated at asymptotic infinity. Hence, the effective complex SYK action is inherently HT-like.

We end this section with a discussion on the choice of the abelian gauge group for the HT sector.

\subsection{Unimodular time at the boundary versus boundary unimodular time}
In section \ref{sec:2}, we have shown that evaluating on-shell the boundary terms for HT$_2$ gravity under mixed boundary conditions for the gauge field leads to the effective action for complex SYK, that is the Schwarzian action coupled to the free particle on $U(1)$:
\begin{equation}\label{eq:quadraticbdyaction}
    I = -\int_{\partial \mathcal{M}} du \, \left( i\Lambda_b\,\widetilde{\varphi}' + \frac{\kappa \epsilon}{2} \Lambda_b^2\, \sqrt{h} \right) \qquad \Longleftrightarrow \qquad I = -\frac{1}{2\kappa} \int_{\partial \mathcal{M}} du \, \widetilde{\varphi}'^2,
\end{equation}
where we added the chemical potential term via the shifted phase field $\widetilde{\varphi} = \varphi - i \mu f$\footnote{The shift does not affect the phase space construction (it is a linear canonical transformation), nor the overall action.}. These two actions are related by performing the Gaussian integral over $\Lambda_b$: The former is written in the phase-space variables $(\widetilde{\varphi},i\Lambda_b)$, which we refer to as the quadratic $\Lambda_b^2$-action; whereas the latter is a purely configuration-space action. We shall not worry about the Schwarzian action throughout this section and focus solely on the free particle sector, where the boundary unimodular time is relevant.

\paragraph{Boundary unimodular time.}

The equivalence between the quadratic and configuration space boundary actions \eqref{eq:quadraticbdyaction} mirrors the on-shell equivalence between 2d Maxwell theory and the $\chi^{2}$-action discussed earlier, now reduced by one spatial dimension in the boundary. This motivates the following field redefinition of the boundary variables $(\widetilde{\varphi},i\,\Lambda_b)$:
\begin{equation}\label{eq:bdyfieldredefinition}
    \Psi(u) = -\frac{\widetilde{\varphi}(u)}{\kappa \epsilon \Lambda_b(u)}, \qquad \qquad  \Theta(u) = - \frac{\kappa \epsilon \Lambda_b^2(u)}{2},
\end{equation}
This field redefinition permits a reformulation of the quadratic action \eqref{eq:quadraticbdyaction}, originally written in terms of $\Lambda_b$, using the variable $\Theta$, which now acts as a Lagrange multiplier. The resulting action becomes linear in $\Theta$ and reads:
\begin{equation}
    I = -\int_0^\beta du \, \Theta \, \bigl( i \Psi' - \sqrt{h} \bigr) - i\,\Theta\,\Psi\Big|_{u=0}^{u=\beta}.
\end{equation}
This is manifestly written as an observer action \cite{Witten:2023xze} and coupled to the Schwarzian mode. Furthermore, this boundary action is also in the form of a one-dimensional Henneaux-Teitelboim action. Therefore, the bulk observer naturally induces a corresponding observer residing at the boundary. In fact, the field $\Theta$ enforces the unimodular constraint on the boundary, namely:
\begin{equation}\label{eq:bdyHTconstraint}
    \sqrt{h} = i\,\Psi',
\end{equation}
and where $\Theta$ is constant on-shell. In fact, the boundary HT constraint \eqref{eq:bdyHTconstraint}, when written in terms of the old boundary variables, translates to $-i \widetilde{\varphi}' = -\kappa \Lambda_b$, which is the mixed boundary condition relating the phase field to the boundary vacuum cosmological constant. 
%\toghether{This further confirms that the boundary action is the correct form.}
Moreover, the resulting theory couples minimally to the Schwarzian mode. Importantly, since the Schwarzian action coupled to a free particle on $U(1)$ reproduces the effective action of the complex SYK model, this identification allows us to interpret that system as a realisation of HT gravity intrinsic to the thermal circle. 

Since we are dealing with a $(0+1)$-dimensional HT theory, let us integrate over the boundary parameter $u$ both sides of \eqref{eq:bdyHTconstraint}. This simply gives an expression for the volume of the thermal circle, or in this case, the length of the boundary:
\begin{equation}
    \int_0^\beta du \, \sqrt{h} = i \int_0^\beta du \, \Psi' \qquad \Longrightarrow \qquad \mathrm{Vol}(S^1_\beta)= i\,\Psi(\beta) - i \Psi(0).
\end{equation}
Here, we are in a position to introduce the very notion of boundary unimodular time:
\begin{equation}\label{eq:bdyunimodulartime}
    T_{\Theta}(u) := i \Psi(u) \qquad \Longrightarrow \qquad \mathrm{Vol}(S^1_\beta) = \Delta T_\Theta = T_{\Theta}(\beta) - T_{\Theta}(0).
\end{equation}
The boundary unimodular time $T_{\Theta}$ is canonically conjugate to $\Theta$ and, on-shell, equals the boundary length. This motivates interpreting $\Theta$ as the intrinsic vacuum cosmological constant of the boundary theory.

It should be distinguished from the quantity $\Lambda_b$, which is not a genuine boundary cosmological constant of the boundary theory, but merely the bulk vacuum cosmological constant $\Lambda$ evaluated at the boundary. In contrast, $\Theta$ is a \textit{bona fide} vacuum cosmological constant intrinsic to the boundary and canonically conjugate to $T_{\Theta}$.
\begin{comment}
\footnote{\toghether{The HT constraint \eqref{eq:bdyHTconstraint} can be rewritten (This can be written in a non densities form) as follows:
\begin{equation}
    1 = i \, \psi' \qquad \psi = -\frac{\widetilde{\varphi}}{\kappa \Lambda_b}.
\end{equation}
This make sure we can define a renormalised boundary unimodular time and volume:
\begin{equation}
    \overline{T}_\Theta = i \psi \qquad \Longrightarrow \qquad \overline{\mathrm{Vol}}(S^1_\beta) = \Delta \overline{T}_\Theta.
\end{equation}}}
\end{comment}

\paragraph{Relating bulk and boundary unimodular times.}

With the $\Lambda_b^2$-action written as an HT gravity theory, and boundary unimodular time $T_\Theta$ defined, consider the configuration space action in \eqref{eq:quadraticbdyaction} which depends only on the shifted phase field. We can then calculate the boundary momentum conjugate to $\widetilde{\varphi}$ to find
\begin{equation}
    p_{\widetilde{\varphi}}(u) = -\frac{\widetilde{\varphi}'}{\kappa} = -i\,\Lambda_b,
\end{equation}
where we used the mixed boundary condition in the form $\widetilde{\varphi}' = i\kappa\,\Lambda_b$ in the last equality. We see that $p_{\widetilde{\varphi}}$ is proportional to the boundary value of the vacuum cosmological constant. By inspection, the same boundary momentum conjugate to $\widetilde{\varphi}$ follows from the quadratic action \eqref{eq:quadraticbdyaction}. However, we can also calculate the momentum conjugate to $\Lambda_b$. This is given explicitly as
\begin{equation}
    p_{\Lambda_b}(u) = -i \widetilde{\varphi}(u) =  -\int^u du \, (A_u - \sqrt{h}).
\end{equation}
The integral expression for $p_{\Lambda_b}$ in fact enables us to express a relation with the bulk unimodular time \eqref{eq:bulkunimodulartime} at the asymptotic infinity: in radial coordinates $(\tau,r)$, we can evaluate it at the constant slice $r = r_\infty$, and find that:
\begin{equation}
    T_\Lambda(r_\infty) = \int_0^\beta du \, A_u = -\Delta p_{\Lambda_b} + \frac{\beta}{\epsilon},
\end{equation}
where $\Delta p_{\Lambda_b} = p_{\Lambda_b}(\beta) - p_{\Lambda_b}(0)$. We see that $T_\Lambda(r_\infty)$ evaluates over the entire boundary circle, and hence, involves the difference in the boundary momentum. 

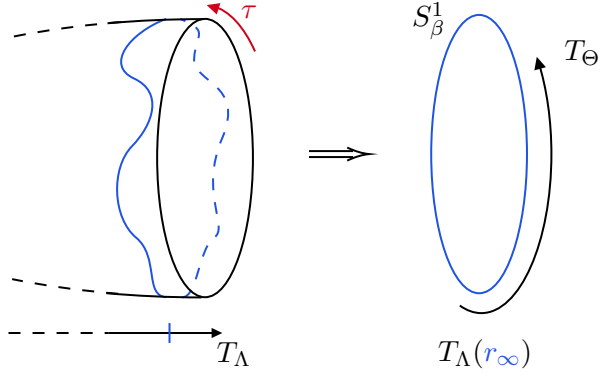
\begin{figure}[!ht]
    \centering

\tikzset{every picture/.style={line width=0.75pt}} %set default line width to 0.75pt        

\begin{tikzpicture}[x=0.75pt,y=0.75pt,yscale=-1,xscale=1]
%uncomment if require: \path (0,208); %set diagram left start at 0, and has height of 208

%Curve Lines [id:da34228143788089893] 
\draw [color={rgb, 255:red, 30; green, 83; blue, 227 }  ,draw opacity=1 ][line width=0.75]    (248.67,152) .. controls (237.67,151.09) and (245.33,131.76) .. (233.67,122) .. controls (222,112.24) and (220.87,87.22) .. (231.67,76) .. controls (242.47,64.78) and (247.27,48.82) .. (232.67,42) .. controls (218.07,35.18) and (230.33,19.09) .. (248.67,12) ;
%Curve Lines [id:da5562729879242112] 
\draw [color={rgb, 255:red, 30; green, 83; blue, 227 }  ,draw opacity=1 ][line width=0.75]  [dash pattern={on 4.5pt off 4.5pt}]  (256.67,152) .. controls (264.33,152.42) and (265,127.09) .. (272.67,122) .. controls (280.33,116.91) and (265,99.76) .. (276.67,72) .. controls (288.33,44.24) and (262.07,44.42) .. (262.67,32) .. controls (263.27,19.58) and (270.87,18.42) .. (255.67,12) ;
%Shape: Arc [id:dp5317147159479556] 
\draw  [draw opacity=0] (268.9,5.05) .. controls (278.4,5.75) and (286.95,14.58) .. (293.06,28.39) -- (267.67,80) -- cycle ; \draw [color={rgb, 255:red, 208; green, 2; blue, 27 }  ,draw opacity=1 ]   (271.86,5.53) .. controls (280.17,7.66) and (287.59,16.03) .. (293.06,28.39) ;  \draw [shift={(268.9,5.05)}, rotate = 17.58] [fill={rgb, 255:red, 208; green, 2; blue, 27 }  ,fill opacity=1 ][line width=0.08]  [draw opacity=0] (6.25,-3) -- (0,0) -- (6.25,3) -- cycle    ;
%Shape: Arc [id:dp012567761030792046] 
\draw  [draw opacity=0] (219.43,14.21) .. controls (234.66,12.84) and (250.59,12.11) .. (267,12.11) .. controls (267.14,12.11) and (267.29,12.11) .. (267.44,12.11) -- (267,82.11) -- cycle ; \draw   (219.43,14.21) .. controls (234.66,12.84) and (250.59,12.11) .. (267,12.11) .. controls (267.14,12.11) and (267.29,12.11) .. (267.44,12.11) ;  
%Shape: Ellipse [id:dp8317397086747285] 
\draw   (268.11,12) .. controls (281.31,12) and (292.01,43.34) .. (292.01,82) .. controls (292.01,120.66) and (281.31,152) .. (268.11,152) .. controls (254.9,152) and (244.2,120.66) .. (244.2,82) .. controls (244.2,43.34) and (254.9,12) .. (268.11,12) -- cycle ;
%Shape: Arc [id:dp882359903467199] 
\draw  [draw opacity=0][dash pattern={on 4.5pt off 4.5pt}][line width=0.75]  (172.7,20.85) .. controls (190.31,17.32) and (209.6,14.74) .. (230.03,13.3) -- (267.67,82) -- cycle ; \draw  [dash pattern={on 4.5pt off 4.5pt}][line width=0.75]  (172.7,20.85) .. controls (190.31,17.32) and (209.6,14.74) .. (230.03,13.3) ;  
%Shape: Arc [id:dp8402550093608042] 
\draw  [draw opacity=0] (266.3,152.11) .. controls (249.94,152.09) and (234.05,151.34) .. (218.87,149.96) -- (267,82.11) -- cycle ; \draw   (266.3,152.11) .. controls (249.94,152.09) and (234.05,151.34) .. (218.87,149.96) ;  
%Shape: Arc [id:dp919942703038006] 
\draw  [draw opacity=0][dash pattern={on 4.5pt off 4.5pt}][line width=0.75]  (224.72,150.27) .. controls (205.91,148.74) and (188.12,146.23) .. (171.81,142.9) -- (268,82) -- cycle ; \draw  [dash pattern={on 4.5pt off 4.5pt}][line width=0.75]  (224.72,150.27) .. controls (205.91,148.74) and (188.12,146.23) .. (171.81,142.9) ;  
%Shape: Ellipse [id:dp8510893260577057] 
\draw  [color={rgb, 255:red, 30; green, 83; blue, 227 }  ,draw opacity=1 ] (404.91,10) .. controls (418.11,10) and (428.81,41.34) .. (428.81,80) .. controls (428.81,118.66) and (418.11,150) .. (404.91,150) .. controls (391.7,150) and (381,118.66) .. (381,80) .. controls (381,41.34) and (391.7,10) .. (404.91,10) -- cycle ;
%Straight Lines [id:da2571778457969609] 
\draw    (220,170) -- (273.74,170) ;
\draw [shift={(276.74,170)}, rotate = 180] [fill={rgb, 255:red, 0; green, 0; blue, 0 }  ][line width=0.08]  [draw opacity=0] (5.36,-2.57) -- (0,0) -- (5.36,2.57) -- cycle    ;
%Shape: Arc [id:dp9456335677727903] 
\draw  [draw opacity=0] (433.7,31.08) .. controls (438.28,44.61) and (441,61.58) .. (441,80) .. controls (441,124.18) and (425.33,160) .. (406,160) .. controls (402.27,160) and (398.67,158.66) .. (395.29,156.19) -- (406,80) -- cycle ; \draw [color={rgb, 255:red, 0; green, 0; blue, 0 }  ,draw opacity=1 ]   (434.63,33.97) .. controls (438.64,46.99) and (441,62.87) .. (441,80) .. controls (441,124.18) and (425.33,160) .. (406,160) .. controls (402.27,160) and (398.67,158.66) .. (395.29,156.19) ;  \draw [shift={(433.7,31.08)}, rotate = 74.37] [fill={rgb, 255:red, 0; green, 0; blue, 0 }  ,fill opacity=1 ][line width=0.08]  [draw opacity=0] (6.25,-3) -- (0,0) -- (6.25,3) -- cycle    ;
%Straight Lines [id:da32489620137582464] 
\draw    (320,78.5) -- (342,78.5)(320,81.5) -- (342,81.5) ;
\draw [shift={(350,80)}, rotate = 180] [color={rgb, 255:red, 0; green, 0; blue, 0 }  ][line width=0.75]    (10.93,-3.29) .. controls (6.95,-1.4) and (3.31,-0.3) .. (0,0) .. controls (3.31,0.3) and (6.95,1.4) .. (10.93,3.29)   ;
%Straight Lines [id:da408527015378749] 
\draw  [dash pattern={on 4.5pt off 4.5pt}]  (170,170) -- (220,170) ;
%Straight Lines [id:da3422525430790082] 
\draw [color={rgb, 255:red, 30; green, 83; blue, 227 }  ,draw opacity=1 ]   (250.37,165) -- (250.37,175) ;

% Text Node
\draw (284.67,4) node [anchor=north west][inner sep=0.75pt]  [color={rgb, 255:red, 208; green, 2; blue, 27 }  ,opacity=1 ]  {$\tau $};
% Text Node
\draw (282.5,180.5) node    {$T_{\Lambda }$};
% Text Node
\draw (446,21.4) node [anchor=north west][inner sep=0.75pt]    {$T_{\Theta }$};
% Text Node
\draw (408,180.5) node   {$T_{\Lambda }(\textcolor[rgb]{0.12,0.33,0.89}{r_\infty })$};
% Text Node
\draw (370,2.0) node [anchor=north west][inner sep=0.75pt]    {$S^{1}_\beta$};

\end{tikzpicture}
    \caption{Here, we focus on the boundary, and the figure depicts how the bulk unimodular time relates to the boundary unimodular time.}
    %\label{fig:placeholder}
\end{figure}

In fact, let us examine the bulk and boundary HT equations and see if there is any relation between them. The bulk HT constraint ensures that the volume of the Euclidean disk is given as a difference of bulk unimodular times, between the horizon $r=r_h$ and asymptotic infinity $r=r_\infty$ constant slices:
\begin{equation}
    \mathrm{Vol}(EAdS_2) = \Delta T_{\Lambda} = T_{\Lambda}(r_\infty) - T_{\Lambda}(r_h).
\end{equation}
Calculating $T_{\Lambda}$ using the gauge field solution on the thermal circle $A_u = \tfrac{1}{\epsilon} + i\widetilde{\varphi}'$, we find that
\begin{equation}
    \Delta T_{\Lambda} = T_\Lambda(r_{\infty}),
\end{equation}
where $T_{\Lambda}(r_h) = 0$ by the regularity and smoothness condition at the horizon with $A_\tau(\tau,r_h) = 0$. Therefore, for the thermal disk topology, the bulk volume only cares about the bulk unimodular time evaluated at the asymptotic infinity.

Now, it is possible to relate $p_{\Lambda_b}$ to $T_{\Theta}$ by substituting \eqref{eq:bdyfieldredefinition} in the expression for the boundary unimodular time \eqref{eq:bdyunimodulartime}, and comparing it with the boundary momentum conjugate to $\Lambda_b$:
\begin{equation}
    T_{\Theta}(u) = \frac{p_{\Lambda_b}(u)}{\kappa \epsilon\Lambda_b}.
\end{equation}
From this, the difference in the boundary unimodular time can be written as 
\begin{equation}\label{eq:bdybulkunimodulartime}
    \Delta T_{\Theta} = \frac{\Delta p_{\Lambda_b}}{\kappa \epsilon\Lambda_b} = - \frac{1}{\kappa \epsilon \Lambda_b} \left(T_{\Lambda}(r_\infty) - \frac{\beta}{\epsilon}\right).
\end{equation}
It is important to emphasise that this relation holds only for the \textit{full} increment of boundary unimodular time, namely when the boundary coordinate runs once around the entire thermal circle, $u\in[0,\beta]$. In this sense, $\Delta T_{\Theta}$ is naturally tied to the bulk unimodular time evaluated at asymptotic infinity: the boundary hypersurface is the \textit{asymptotic} slice, and it wraps the full thermal circle on which the bulk unimodular time is to be read off. To our knowledge, this is the first instance in which unimodular time is defined in Euclidean (AdS) spacetime and a precise relation is established between two notions of unimodular time, namely the bulk and boundary constructions.

Furthermore, from \eqref{eq:bdybulkunimodulartime}, and the bulk/boundary HT constraint equations, we find a relation between bulk and boundary volumes:
\begin{equation}
    \mathrm{Vol}(S^{1}_\beta) = - \frac{1}{\kappa \epsilon \Lambda_b} \left(\mathrm{Vol}(EAdS_2) - \frac{\beta}{\epsilon} \right).
\end{equation}
The right hand side is simply a renormalised bulk volume $\overline{\mathrm{Vol}}(EAdS_2)$, and therefore, we have a very simple and elegant expression relating bulk and boundary volumes:
\begin{equation}
    \overline{\mathrm{Vol}}(EAdS_2) = -\kappa \epsilon \Lambda_b \mathrm{Vol}(S^1_\beta).
\end{equation}
This gives a rather interesting equality on the thermal circle. The left hand side gives $\overline{\mathrm{Vol}}(EAdS_2) = -\beta r_h \equiv \beta \mu$. However, the right hand side is simply $-\kappa \epsilon \Lambda_b \mathrm{Vol}(S^1_\beta) = -\kappa \beta \Lambda_b$. Therefore, we find that
\begin{equation}\label{eq:chempot=bdyvac}
    \mu = -\kappa \Lambda_b.
\end{equation}
This expression is compatible with the mixed boundary conditions and, by equivalence, with the boundary HT constraint. In particular, the condition that identifies the boundary volume with the difference of boundary unimodular times reduces here to the simple relation that the chemical potential equals the boundary vacuum cosmological constant. Moreover, from \eqref{eq:chempot=bdyvac} written with $\mu = -r_h$:
\begin{equation}
    r_h=\kappa\,\Lambda_b,
\end{equation}
we see that the vacuum cosmological constant evaluated at the boundary acquires an explicit temperature dependence through $r_h$ (recall that $r_h \propto \beta^{-1}$). This feature is reminiscent of the behaviour encountered in the $\reallywidehat{\text{CGHS}}$ model \cite{Afshar:2019axx,Kar:2022vqy,Kar:2022sdc}.

\paragraph{Fermion number $\mathcal{Q}$ as $\Lambda_b$.}

Now that we have a well understanding for the boundary unimodular time, we turn to give meaning to the boundary cosmological constant $\Lambda_b$ in terms of complex SYK language. Going back to the configuration space action \eqref{eq:quadraticbdyaction}, we find that $p_{\widetilde{\varphi}}$ is conserved. This ensures that, by virtue of Noether's theorem, we have a conserved Noether charge
\begin{equation}
    \mathcal{Q} = p_{\widetilde{\varphi}}(u) = -\frac{\widetilde{\varphi}'}{\kappa}.
\end{equation}
Equivalently, from $p_{\widetilde{\varphi}} = -i \Lambda_b$ taken constant on-shell, we have that
\begin{equation}
    \mathcal{Q} = -i \Lambda_b.
\end{equation}
This is once again not only consistent with the fact that $\mu = -\kappa \Lambda_b$, but as well the boundary HT constraint equation. Indeed, the boundary charge $\mathcal{Q}$ is a manifestation of the invariance of the free particle action under constant translations $\widetilde{\varphi} \rightarrow \widetilde{\varphi} + \widetilde{\varphi}_0$. Within this context, the boundary charge $\mathcal{Q}$ corresponds to the fermion number of the complex SYK model \cite{Davison:2016ngz}. Therefore, $\Lambda_b$ is the fermion number or $U(1)$ charge in the dual complex SYK theory.

\subsection{A discussion on compact and non-compact groups}

In this subsection, we explain why our results do not depend on the specific choice of abelian gauge group, and in particular why working with $\mathbb{R}$ instead of $U(1)$ leaves all conclusions of this paper unchanged.

We can start our discussion by noting that, from the previous subsection, the volume of the thermal disk can be seen as:  
\begin{equation}
    \mathrm{Vol}(EAdS_2) = \beta (r_\infty - r_h) \qquad \text{and} \qquad \mathrm{Vol}(EAdS_2) \equiv T_{\Lambda}(r_\infty),
\end{equation}
where one reproduces the volume by computing $T_\Lambda(r_\infty)$ satisfying the unimodular condition, which involves a choice for the value of the difference of the phase field $\Delta \varphi$. In particular, on $S^1_\beta$, we impose the following conditions on the phase mode and the time-reparametrisation field:
\begin{equation}
    \varphi(u+\beta) - \varphi(u) \in 2 \pi \mathbb{Z}, \ \ \ \ \ \ \ \ \ \ \ \ \ \ f(u+\beta) = f(u) + \beta.
\end{equation}
Therefore, in terms of the periodicity condition on the boundary momentum, we get:
\begin{equation}\label{eq:bulkUTatinfty}
    T_{\Lambda}(r_\infty) \equiv -\Delta p_{\Lambda_b}  + \frac{\beta}{\epsilon}= \beta(r_\infty - r_h) + 2 \pi i n, \qquad n \in \mathbb{Z}.
\end{equation}
One may then ask whether the unimodular time evaluated at the boundary should admit a winding contribution\footnote{Indeed, for a generic choice of bulk slicing, nontrivial winding can in principle appear, reflecting the possibility that the phase mode acquires a net monodromy around the circle.}. The computation of the volume should, in principle, not give any winding for the Euclidean disk. Requiring this volume to be on-shell equivalent to the unimodular time enforces the vanishing of the phase mismatch, namely $\Delta \varphi = 0$. Therefore, consistency with the equations of motion of HT$_2$ gravity requires the phase field to be periodic on the thermal disk topology:
\begin{equation}
    \varphi(u+\beta) = \varphi(u).
\end{equation}

Setting $n=0$ might seem to suggest that the relevant gauge group is $\mathbb{R}$ rather than $U(1)$; however, this conclusion is unwarranted. The point is that winding is a phenomenon associated with loops that wrap the thermal circle. In the Euclidean black-hole geometry, the thermal circle is contractible in the bulk: since it bounds the Euclidean disk, a loop at the cut-off surface can be continuously shrunk through the interior and collapsed at the smooth tip of the cigar (the horizon at $r=r_h$). Regularity at the tip, therefore, enforces trivial holonomy and, equivalently, periodicity of the phase field. This requirement is thus dictated by the topology of the underlying geometry rather than by a change of gauge group.

\paragraph{Particle on $\mathbb{R}$ is not an issue.}

The only two abelian groups relevant in the HT gravity context are $U(1)$ and $\mathbb{R}$. The former, which we have used throughout this paper, admits both little and large gauge transformations. More concretely, little transformations are single-valued in $\varphi$, i.e. $\varphi(u+\beta)=\varphi(u)$, whereas large transformations involve winding, so that $\varphi$ becomes multi-valued:
\begin{equation}
    \varphi(u+\beta)=\varphi(u)+2\pi n,\qquad n\in\mathbb{Z}.
\end{equation}
If instead one takes the gauge group to be $\mathbb{R}$, only little gauge transformations are possible. This follows from the fact that all $\mathbb{R}$-bundles are trivial over Riemann surfaces. In particular, one could take $\mathbb{R}$ as the gauge group for the HT sector, which would naturally imply that the bulk unimodular time should not contain any winding-type contributions, as seen in \eqref{eq:bulkUTatinfty}. 

Taking $\mathbb{R}$ also does not prevent one from having the free particle action \eqref{eq:quadraticbdyaction}; the only change is the replacement of $U(1)$ by $\mathbb{R}$, with $\varphi$ now a single-valued field. Moreover, from the BF perspective on HT$_2$ gravity and its central extension, the Lie algebras of $\mathbb{R}$ and $U(1)$ coincide, so the corresponding results are likewise unaffected. Finally, one may read the entire paper with gauge group $\mathbb{R}$ in mind rather than $U(1)$, and all results obtained up to this point remain intact.

\section{Outlook}

Henneaux-Teitelboim unimodular gravity has appeared in a variety of settings, ranging from four-dimensional constructions to two-dimensional models. In this work, we pursued a different direction by focusing on its Euclidean formulation and by sharpening its holographic interpretation, a viewpoint that is largely absent from the standard HT gravity literature. Working in Euclidean $AdS_2$ allowed us to make the boundary dynamics and the role of unimodular time completely explicit. We finish this paper by mentioning the main results and remaining open problems to be addressed in future work:
\begin{itemize}
    \item By imposing consistent mixed boundary conditions and the accompanying boundary terms, we showed that Euclidean HT$_2$ in $AdS_2$ reproduces at the boundary the same universal degrees of freedom as the low-energy effective theory of complex SYK: the Schwarzian coupled to a free particle on $U(1)$. A natural next step is to better understand the UV completion of this duality. It would also be interesting to explore higher-dimensional HT theories with appropriately chosen boundary terms, with the aim of isolating an effective boundary sector that plays the role of a higher-dimensional analogue of the particle action discussed throughout the paper.

    \item We make explicit that HT$_2$ gravity can be rewritten as JT gravity coupled to a suitable 2d Maxwell sector. In particular, the emergence of the particle on $U(1)$ is most naturally understood only after correctly accounting for the boundary terms and constraints in the Maxwell description: the relevant bulk physics is closer to a top form (flux) sector controlling the vacuum cosmological constant than to a fully dynamical, propagating Maxwell field, which clarifies why seemingly different bulk actions can nevertheless reduce to the same boundary $U(1)$ free-particle theory. More broadly, this suggests a guiding principle for generalisations: whenever the vacuum energy is effectively governed by higher-form flux, an HT-like rewriting should be available, potentially isolating a universal boundary mode in holographic settings. A concrete arena where this expectation is natural is higher-dimensional supergravity, where flux compactifications generate $AdS$ vacua supported by top form field strengths. For example, the near-horizon geometries $AdS_5\times S^5$ of D3-branes in type IIB and $AdS_4\times S^7$ of M2-branes in M-theory (supported by the four-form) suggest that, after consistent truncation to the $AdS$ sector, the resulting lower-dimensional theory may admit an HT-like reformulation in which the cosmological constant is traded for a top form field strength.
    
    \item We connected HT$_2$ to a BF-type gauge theory and clarified how the relevant symmetry algebra controls the coupling between the JT and $U(1)$ sectors, and also highlighted the distinction between the non flat and flat limits. We also found the existence of a minimally extended HT theory when embedding HT$_2$ within higher-dimensional black hole physics. It is likely important to determine the correct boundary description of the resulting extended theory, with existing insights in related contexts \cite{Iliesiu:2019lfc} suggesting that such embeddings could yield interesting boundary theories. 
    
    \item At the boundary, we exhibited a $(0+1)$-dimensional HT theory obtained by field redefinition of the $U(1)$ particle action, which naturally introduces a boundary unimodular time conjugate to the cosmological constant intrinsic to the boundary. This makes it possible to reinterpret the particle on a group action as an observer action: it provides a natural observer for the Schwarzian mode and promotes the phase field to a boundary clock. In this language, the boundary unimodular time is encoded in the phase mode and the chemical potential. It would be particularly interesting to study boundary unimodular time on more general $AdS_2$ backgrounds, such as Euclidean wormholes with arbitrary genus and boundaries. In this setting, analysing the HT$_2$ gravitational path integral with a sum over topologies may shed light on topology change in 2d Euclidean quantum gravity, and in particular on whether unimodular time furnishes a natural organising variable for such processes.
    
    \item Finally, it would be relevant to apply the holographic technology developed in this paper to de Sitter spacetime, where Schwarzian$+U(1)$-type boundary dynamics at asymptotic future/past infinity $\mathcal{I}^\pm$ may be recast as a $(0+1)$-dimensional HT system. This would endow $\mathcal{I}^\pm$ with a natural boundary time and a canonical quantisation of the boundary Hilbert space, potentially alleviating the longstanding tension between bulk and boundary quantisation in de Sitter. In particular, one may ask whether unimodular time can provide a clean notion of evolution and a sharper construction of holographic observables in de Sitter, thereby completing the picture initiated in our previous work on the bulk sector of de Sitter HT$_2$ \cite{Alexandre:2025rgx}.
\end{itemize}

%---------------- Acknowledgments --------------------------
\acknowledgments

We are grateful to Andreas Blommaert for numerous conversations that were integral to this work. We also thank Hamid R. Afshar, Adri\'{a}n S\'{a}nchez-Garrido, Steffen Gielen, Oliver Gould, Felix M. Haehl, Jo\~{a}o Magueijo, Ioannis Matthaiakakis, Benjamin Muntz, Antonio Padilla, Moshe Rozali, Bayram Tekin, and Bruno G. Umbert for helpful discussions. AE is supported by UK Research and Innovation (UKRI) under the UK government’s Horizon Europe Funding Guarantee (EP/X030334/1). FSR is supported by the Royal Society Dorothy Hodgkin Fellowship.

%---------------- Appendix --------------------------

\appendix

\section{The original unimodular gravity}\label{app:A}

We provide a brief overview of the standard framework of unimodular gravity.

In 1919, in an attempt to solve his field equations, Einstein introduced the gauge fixing $\sqrt{-g} = 1$ \cite{Einstein-unimod}. This statement amounts to the observation that tensor densities can be treated on essentially the same footing as ordinary tensors. With this condition, the action (without a cosmological constant) becomes
\begin{equation}
    S = \frac{1}{2} \int_{\mathcal{M}} d^4x \,\sqrt{-g} \,  R \, \, \, \, \, \, \Longrightarrow \, \, \, \, \, \, S = \frac{1}{2} \int_{\mathcal{M}} d^4x \, R.
\end{equation}
Obviously, this action is not invariant under the full spacetime diffeomorphisms, but only under volume-preserving coordinate transformations. With the determinant of the metric fixed, we can assume that at the level of the variation of the action, one can take
\begin{equation}
    \delta \sqrt{-g} = 0 \, \, \, \, \, \, \Longleftrightarrow \, \, \, \, \, \, g_{\mu \nu} \delta g^{\mu \nu} = 0.
\end{equation}
The last condition says that we consider traceless variations of the metric tensor in the action. Dropping, for the sake of simplicity, various boundary terms, the variation of the action under traceless variations of the metric gives
\begin{equation}
    0 = \delta S = \frac{1}{2} \int_{\mathcal{M}} d^4x \, R_{\mu \nu} \delta g^{\mu \nu} \equiv \frac{1}{2} \int_{\mathcal{M}} d^4x \, G_{\mu \nu} \delta g^{\mu \nu},
\end{equation}
where we used in the last equality the fact that $G_{\mu \nu} \delta g^{\mu \nu} = R_{\mu \nu} \delta g^{\mu \nu} - \tfrac{1}{2} R \,g_{\mu \nu} \delta g^{\mu \nu} = R_{\mu \nu} \delta g^{\mu \nu}$, as the second term vanishes due to traceless variations of the metric tensor. Now, $\delta g^{\mu \nu}$ being not arbitrary and determined up to it being traceless, we can use the fact that for a symmetric rank-2 tensor $G_{\mu \nu}$, if it is subject to $g_{\mu \nu} \delta g^{\mu \nu} = 0$, then:
\begin{equation}
    \int_{\mathcal{M}} d^4x \, G_{\mu \nu} \delta g^{\mu \nu} = 0 \, \, \, \, \, \, \, \Longleftrightarrow \, \, \, \, \, \, G_{\mu \nu} \, \, \, \text{is pure trace}.
\end{equation}
The fact that $G_{\mu \nu}$ is pure trace means that we only get its traceless part, i.e.
\begin{equation}
    G_{\mu \nu} \, \, \, \text{is pure trace} \, \, \, \, \, \Longleftrightarrow \, \, \, \, \, G^{\text{TF}}_{\mu \nu}  \equiv G_{\mu \nu} - \frac{1}{4} g_{\mu \nu} G = 0.
\end{equation}
This can be obtained from the projector tensor:
\begin{equation}
    \Pi_{\mu \nu \,}^{\ \ \, \alpha \beta} := \delta_{\mu}^{\, \, (\alpha} \delta_{\nu}^{\, \, \beta)} - \frac{1}{4}g_{\mu \nu} g^{\alpha \beta},
\end{equation}
such that $G^{\text{TF}}_{\mu \nu} = \Pi_{\mu \nu \,}^{\ \ \, \alpha \beta} G_{\alpha \beta}$. Rather importantly, the variation of the metric is invariant under the projector map:
\begin{equation}
    \Pi_{\mu \nu}^{\ \ \ \alpha \beta} \delta g^{\mu \nu} = \delta g^{\alpha \beta}.
\end{equation}
In fact, we find that the field equations of unimodular gravity are given by the traceless Einstein equations:
\begin{equation}
    R_{\mu \nu} - \frac{1}{4} g_{\mu \nu} R = 0.
\end{equation}

We can now use the Bianchi identities $\nabla^{\mu} R_{\mu \nu} = \frac{1}{2} g_{\mu \nu} \nabla^{\mu}R$ and take the derivative of the traceless equations to find:
\begin{equation}
    \partial^{\mu}R = 0 \, \, \, \, \,  \Longrightarrow \, \, \, \, \, R = R_0 = \text{const.}
\end{equation}
Substituting this into the traceless equations with $G = -R = -R_0$ gives the cosmological constant as an integration constant:
\begin{equation}
    0 = G_{\mu \nu} + \frac{1}{4} g_{\mu \nu} G = G_{\mu \nu} +  \Lambda_0 g_{\mu\nu} \, \, \, \, \, \, \, \, \, \, \text{where \, \, \, \, } R_0 = 4 \Lambda_0.
\end{equation}
Therefore, we find that the traceless equations reduce to Einstein equations with a cosmological constant given as an integration constant.

It is rather remarkable that unimodular gravity, although invariant under volume-preserving diffeomorphisms, still provides the same dynamics as classical GR. We would like to point out that there is an alternative action that gives similar results, namely the one where we do not gauge fix $\sqrt{-g} = 1$, but nevertheless demand traceless variations of the metric tensor. Hence, let us consider the fully diffeomorphism-invariant standard Einstein-Hilbert action without a cosmological constant:
\begin{equation}
    S = \frac{1}{2} \int_{\mathcal{M}} d^4x \, \sqrt{-g} \, R,
\end{equation}
subject to $g_{\mu \nu} \delta g^{\mu \nu} = 0$. It results that the variation gives (up to possible boundary terms that we ignore)
\begin{equation}
    0 = \delta S = \frac{1}{2} \int_{\mathcal{M}} d^4x \, \sqrt{-g}\, R_{\mu \nu} \delta g^{\mu \nu} = \frac{1}{2} \int_{\mathcal{M}} d^4x \, \sqrt{-g} \, G_{\mu \nu} \delta g^{\mu \nu}.
\end{equation}
Unsurprisingly, this leads again to the traceless Einstein equations and, upon invoking the contracted Bianchi identities, to the full Einstein equations with the cosmological constant emerging as an integration constant. The key difference here is that we have not imposed $\sqrt{-g}=1$; instead, restricting to traceless metric variations effectively encodes the unimodular constraint.

We may wish to recover the full Einstein field equations rather than first projecting onto their traceless form. This can be done by imposing the unimodular constraint $\sqrt{-g} = 1$ at the level of the action with a non-dynamical Lagrange multiplier $\Lambda$, while allowing unrestricted metric variations. The resulting action is invariant only under volume-preserving diffeomorphisms and reads:
\begin{equation}
    S = \frac{1}{2} \int_{\mathcal{M}} d^4x \, \bigl(\sqrt{-g} \, R - 2 \Lambda (\sqrt{-g} - 1)\bigr).
\end{equation}
Variations with respect to the Lagrange multiplier field lead to $\sqrt{-g} = 1$, and variations with respect to the metric tensor give the standard Einstein equations with a varying cosmological constant, namely
\begin{equation}
    G_{\mu \nu} + \Lambda(x) g_{\mu \nu} = 0.
\end{equation}
Taking the trace leads to the Ricci scalar being proportional to the Lagrange multiplier, i.e. $R = 4 \Lambda(x)$. A simple substitution back into Einstein equations leads to the traceless equations. Then, via Bianchi identities, it is easy to show that $\Lambda$ is an integration constant $\Lambda = \Lambda_{\text{0}}$. Therefore, we find exactly vacuum Einstein field equations with a ``constant'' cosmological constant.

\section{Other boundary conditions}

For completeness, we provide a discussion of other possible boundary conditions for the HT sector. This is complementary to section \ref{sec:2}, where we discussed the mixed boundary conditions for holographic purposes. In particular, one may fix either $\Lambda_b$ or $\mathcal{T}^{\mu}$ at the boundary. Note, however, that under either choice, the on-shell action does not reduce to the free particle on $U(1)$.

We first point out an alternative bulk Euclidean action equivalent to $I_{\text{bulk}}$, up to a possible boundary term, exists and is given by:
\begin{equation}\label{eq:HTprime}
    \widetilde{I}_{\text{bulk}} = -\frac{1}{2} \int_{\mathcal{M}} d^2x \, \sqrt{g}\bigl(\phi(R +2) - 2 \Lambda\bigr) + \int_{\mathcal{M}} d^2x \, \mathcal{T}^{\mu} \partial_{\mu} \Lambda.
\end{equation}
This alternative form of the bulk action will prove useful when we discuss boundary conditions shortly.

\paragraph{Fixing $\Lambda_b$ at the boundary.} 

We first turn to the case of fixed $\Lambda_b$. Because the variation of $I_{\text{bulk}}$ in \eqref{eq:bulkHT2action} produces no term proportional to $\delta \Lambda_b$, an additional boundary contribution is required to cancel \eqref{eq:HTboundary} exactly. Accordingly, the GHY boundary term must be supplemented by:
\begin{equation}\label{eq:fixedlLambdaaction}
    I^{\Lambda_b\text{-fixed}} =  I_{\text{bulk}}- \int_{\partial \mathcal{M}} du\, \sqrt{h} \, \phi_b (K-1) + \int_{\partial \mathcal{M}} du\, n_{\mu} \Lambda_b \mathcal{T}^{\mu}.
\end{equation}
With the equations of motion solved and the asymptotic boundary conditions imposed, we can integrate out both the dilaton and the vacuum cosmological constant. Substituting the corresponding on-shell solutions back into the action then yields the boundary on-shell effective action:
\begin{align}\label{eq:fixedlambdaaction}
    I^{\Lambda_b\text{-fixed}}\Big|_{\text{on-shell}} &= - \int_{\partial \mathcal{M}} du \, \sqrt{h} \, \phi_b (K-1) + \int_{\partial \mathcal{M}} du \, \Lambda_b A_u \nonumber\\
    & = - \overline{\phi}_r \int_{\partial \mathcal{M}} du \, \text{Sch}\{t,u\}  + \frac{\Lambda_b \ell}{\epsilon} +i \Lambda_b\Delta\varphi.
\end{align}
where $\ell = \int_{\partial \mathcal{M}} du$ denotes the boundary length, and $\Delta\varphi$ is the boundary value of $\varphi$. Clearly, $\tfrac{\Lambda_b \ell}{\epsilon}$ is divergent as we take $\epsilon$ to zero. This can be removed by adding a holographic counterterm at the boundary:
\begin{equation}\label{eq:countertermfixedlambda}
    I_{\text{bdy ct}} = -\int_{\partial \mathcal{M}} du \, \sqrt{h} \, \Lambda_b.
\end{equation}
Combining \eqref{eq:fixedlambdaaction} and \eqref{eq:countertermfixedlambda}, the resulting finite on-shell action at the boundary is therefore given by
\begin{equation}\label{eq:fixedLambdaonshellaction}
    I^{\Lambda_b\text{-fixed}}\Big|_{\text{on-shell}} = - \overline{\phi}_r \int_{\partial \mathcal{M}} du \, \, \text{Sch}\{t,u\} + i \Lambda_b \Delta \varphi.
\end{equation}
Therefore, the resulting boundary theory is just the Schwarzian and a possible constant shift. 
%It should be noted that $\Delta \varphi$ is highly dependent on the domain of integration. For instance, in Poincar\'{e} coordinates, we have $\partial \mathcal{M} \cong \mathbb{R}$ and the set of gauge transformations $\mathcal{G} \cong \mathrm{Maps}(\mathbb{R}\rightarrow U(1))$ all have trivial holonomy. This makes $\varphi(\pm \infty) = 0$ and hence $\Delta\varphi = 0$. Consequently, the Schwarzian action provides the sole boundary contribution. For a closed boundary $\partial \mathcal{M} \cong S^1$, gauge transformations may carry a non-trivial holonomy around the circle, leading to a shift $\Delta\varphi = 2\pi n$ with integer $n$. This feature will play an important role in the next section, where we consider the finite-temperature case with the boundary given by the thermal circle $S^1_\beta$.

Before concluding, we note that the alternative action \eqref{eq:HTprime} avoids the need for an additional boundary term for the unimodular sector when $\Lambda_b$ is held fixed. This is because the boundary term arising from the variation is proportional to $\delta \Lambda_b$, which vanishes when $\Lambda_b$ is fixed, leaving only the GHY boundary term:
\begin{align}
    \widetilde{I}^{\Lambda_b\text{-fixed}} = \widetilde{I}_{\text{bulk}}- \int_{\partial \mathcal{M}} du \, \sqrt{h} \, \phi_b (K - 1).
\end{align}
An important remark is that $\widetilde{I}^{\Lambda_b}$ and $I^{\Lambda_b}$ are actually equal up to integration by parts.

\paragraph{Fixing $\mathcal{T}^{\mu}$ at the boundary.}

For the case of fixed $\mathcal{T}^{\mu}$ at the boundary, which allows fluctuating $\Lambda_b$ at the boundary, we have the opposite behaviour compared to the last case. By demanding $\delta \mathcal{T}^{\mu} \big|_{\partial \mathcal{M}} = 0$,  we don't have to add any further boundary terms, and the total action becomes
\begin{align}
    I^{\mathcal{T}^\mu\text{-fixed}} = I_{\text{bulk}} - \int_{\partial \mathcal{M}} du \, \sqrt{h} \, \phi_b (K - 1).
\end{align}
It is rather interesting that fixing $\mathcal{T}^{\mu}$ at the boundary does not seem to change the boundary dynamics. The resultant on-shell action gives a boundary theory that is solely described by the Schwarzian degrees of freedom, as no additional boundary terms are needed to make the variational principle well-defined. From the perspective of $\widetilde{I}_{\text{bulk}}$, the variation reads 
\begin{equation}
    \delta \widetilde{I}_{\text{bulk}} \supset \int_{\mathcal{M}} d^2x \, \sqrt{g} \,\delta \Lambda + \int_{\mathcal{M}} d^2x \, \delta\mathcal{T}^{\mu}( \partial_{\mu} \Lambda) + \int_{\mathcal{M}} d^2x \, \mathcal{T}^{\mu}( \partial_{\mu} \delta \Lambda).
\end{equation}
Integrating the last term by parts yields the boundary contribution
\begin{align}
    \delta \widetilde{I}_{\text{bulk}} \supset &\int_{\mathcal{M}} d^2x \, \sqrt{g} \,\delta \Lambda + \int_{\mathcal{M}} d^2x \, \delta\mathcal{T}^{\mu}( \partial_{\mu} \Lambda) - \int_{\mathcal{M}} d^2x \, \delta \Lambda(\partial_{\mu}\mathcal{T}^{\mu} )\nonumber\\
    &+ \int_{\partial \mathcal{M}} du \, n_{\mu} \, \delta \Lambda_b(u) \mathcal{T}^{\mu}.
\end{align}
As the last term does not vanish, one needs to add a suitable boundary term to the action, leading to the total action:
\begin{align}
    \widetilde{I}^{\mathcal{T}^\mu}_{\text{bdy}} = - \int_{\partial \mathcal{M}} du\, \sqrt{h} \, \phi_b (K-1) - \int_{\partial \mathcal{M}} du\, n_{\mu} \Lambda_b(u) \mathcal{T}^{\mu}.
\end{align}

\vspace{10pt}
\bibliographystyle{JHEP}
%\newpage
\bibliography{main}

\end{document}